\documentclass[journal]{IEEEtran}
 \pdfoutput=1
 \usepackage{cite}
\usepackage{amsmath,amssymb,amsfonts}
\usepackage{algorithmic}
\usepackage{graphicx}
\usepackage{textcomp}
\usepackage{framed,multirow}
\usepackage{latexsym}
\usepackage{amsmath}
\usepackage{threeparttable}
\usepackage{float}
\usepackage{array}
\usepackage{hhline}
\usepackage{colortbl}
\usepackage{stfloats}
\usepackage{booktabs}
\usepackage{multirow}
\usepackage{enumerate}
\usepackage{diagbox}
\usepackage{url}
\usepackage{hyperref}
\usepackage{makecell,rotating}
\usepackage{caption}
\usepackage{color}
\usepackage{arydshln}
\usepackage[linesnumbered, ruled]{algorithm2e}
\DeclareUnicodeCharacter{2061}{}

\DeclareGraphicsExtensions{.pdf,.jpeg,.png,.jpg}

\graphicspath{{fig/}}

\definecolor{mygray}{gray}{.9}
\newcommand{\red}[1]{\textcolor{red}{#1}}

\definecolor{newcolor}{rgb}{.8,.349,.1}

\makeatletter

\newcommand{\Rmnum}[1]{\expandafter\@slowromancap\romannumeral #1@}
\makeatother

\title{FedDBL: Communication and Data Efficient Federated Deep-Broad Learning for Histopathological Tissue Classification}

\author{Tianpeng Deng$^\dagger$, Yanqi Huang$^\dagger$, Zhenwei Shi$^\dagger$, Jiatai Lin, Qi Dou, Zaiyi Liu, Xiao-jing Guo, C. L. Philip Chen, \IEEEmembership{Fellow,~IEEE,} Chu Han
\thanks{This work was supported by Key-Area Research and Development Program of Guangdong Province (No. 2021B0101420006), 
Regional Innovation and Development Joint Fund of National Natural Science Foundation of China (No. U22A20345), 
National Science Foundation for Young Scientists of China (No. 62102103),
Natural Science Foundation for Distinguished Young Scholars of Guangdong Province (No. 2023B1515020043). 
\textit{(Equal contribution: Tianpeng Deng, Yanqi Huang and Guoqiang Han.)}
\textit{(Corresponding author: Xiao-jing Guo, C. L. Philip Chen and Chu Han.)}}
\thanks{Tianpeng Deng, Jiatai Lin, Guoqiang Han and C. L. P. Chen are with the School of Computer Science and Engineering, South China University of Technology, Guangzhou, Guangdong, 510006, China.}
\thanks{Yanqi Huang, Zhenwei Shi, Zaiyi Liu and Chu Han are with the Department of Radiology, Guangdong Provincial People's Hospital (Guangdong Academy of Medical Sciences), Southern Medical University, Guangzhou 510080, China and Guangdong Provincial Key Laboratory of Artificial Intelligence in Medical Image Analysis and Application, Guangzhou 510080, China.}
\thanks{Qi Dou is with the Department of Computer Science and Engineering, The Chinese University of Hong Kong, Hong Kong, 999077, China.}
\thanks{Xiao-jing Guo is with the Department of Breast Pathology and Lab, Tianjin Medical University Cancer Institute and Hospital, National Clinical Research Center of Cancer, Key Laboratory of Cancer Prevention and Therapy, Tianjin, Tianjin's Clinical Research Center for Cancer, Key Laboratory of Breast Cancer Prevention and Therapy, Tianjin Medical University, Ministry of Education, Tianjin 300060, China.}
}

\begin{document}
\maketitle

\IEEEtitleabstractindextext{\begin{abstract}
Histopathological tissue classification is a fundamental task in computational pathology. Deep learning-based models have achieved superior performance but centralized training with data centralization suffers from the privacy leakage problem. Federated learning (FL) can safeguard privacy by keeping training samples locally, but existing FL-based frameworks require a large number of well-annotated training samples and numerous rounds of communication which hinder their practicability in the real-world clinical scenario. In this paper, we propose a universal and lightweight federated learning framework, named Federated Deep-Broad Learning (FedDBL), to achieve superior classification performance with limited training samples and only one-round communication. By simply associating a pre-trained deep learning feature extractor, a fast and lightweight broad learning inference system and a classical federated aggregation approach, FedDBL can dramatically reduce data dependency and improve communication efficiency. Five-fold cross-validation demonstrates that FedDBL greatly outperforms the competitors with only one-round communication and limited training samples, while it even achieves comparable performance with the ones under multiple-round communications. Furthermore, due to the lightweight design and one-round communication, FedDBL reduces the communication burden from 4.6GB to only 276.5KB per client using the ResNet-50 backbone at 50-round training. Since no data or deep model sharing across different clients, the privacy issue is well-solved and the model security is guaranteed with no model inversion attack risk. Code is available at \url{https://github.com/tianpeng-deng/FedDBL}.
\end{abstract}
\begin{IEEEkeywords}
Federated Learning, Data and Communication Efficiency, Deep-Broad Learning, Histopathological Tissue Classification
\end{IEEEkeywords}}

\maketitle
\IEEEdisplaynontitleabstractindextext

\IEEEpeerreviewmaketitle

\section{Introduction}
\IEEEPARstart{T}{issue} classification~\cite{gurcan2009histopathological,FUCHS2011515}, also known as tissue phenotyping, aims to use computer algorithms to automatically recognize different tissue types in the Whole Slide Images (WSIs). It is one of the fundamental tasks in computational pathology~\cite{srinidhi2021deep,wang2019weakly} which can parse the landscape of tumor microenvironment for precise predictions of cancer diagnosis~\cite{bulten2022artificial}, prognosis~\cite{fu2020pan,pages2018international} and treatment response~\cite{vanguri2022multimodal}. With the advancements of deep learning algorithms and the growing number of open data~\cite{kather2019predicting,zhao2020artificial,yang2022two}, this problem has been well studied with outstanding classification performance~\cite{hatami2021deep}. In clinical practice, however, it still faces ethical, regulatory and legal obstacles where centralized data collection may lead to privacy leakage, especially the RAW data. 

Federated Learning (FL)~\cite{yang2019federated,le2021federated} framework provides a promising solution to protect user privacy by only sharing the intermediate results or the model parameters instead of the raw data, which has been widely studied in medical image analysis~\cite{pati2022federated,sheller2020federated}. But only very few attempts~\cite{saldanha2022swarm,shen2022tmi,ke2021isbi} have been made in computational pathology and the research progress still lags behind other medical image modalities~\cite{rauniyar2022federated} due to the following two obstacles.

The first one is the data dependency problem. Since most of the existing FL frameworks are constructed based on deep learning models. They are data-hungry and commonly require a large amount of well-annotated samples. However, labeling histopathological images is time-consuming, expertise-dependent and expensive~\cite{greenwald2022whole,pati2021reducing}. When without enough training samples, existing models may not achieve favorable performance. 
Another obstacle is the communication overhead. The training procedure of traditional FL models needs multiple cloud-client iterations to achieve global convergence. However, deep learning models are with tens of millions of parameters, which greatly increases the communication burden when with multiple communication rounds. 
Although some recent works use self-supervised learning {SSL} methods with unlabelled data to reduce the demand of labelled data \cite{yan2023label,kassem2022federated}, they still need multiple communication rounds to train a stable domain-specific pre-trained model. Lack of training samples may further amplify the communication burden because deep learning models commonly require more iterations to converge when with limited training samples. Moreover, frequent communications may increase the chance of being attacked, such as man-in-the-middle attacks~\cite{wang2020man}.

\begin{figure}[t!]
	\includegraphics[width=\linewidth]{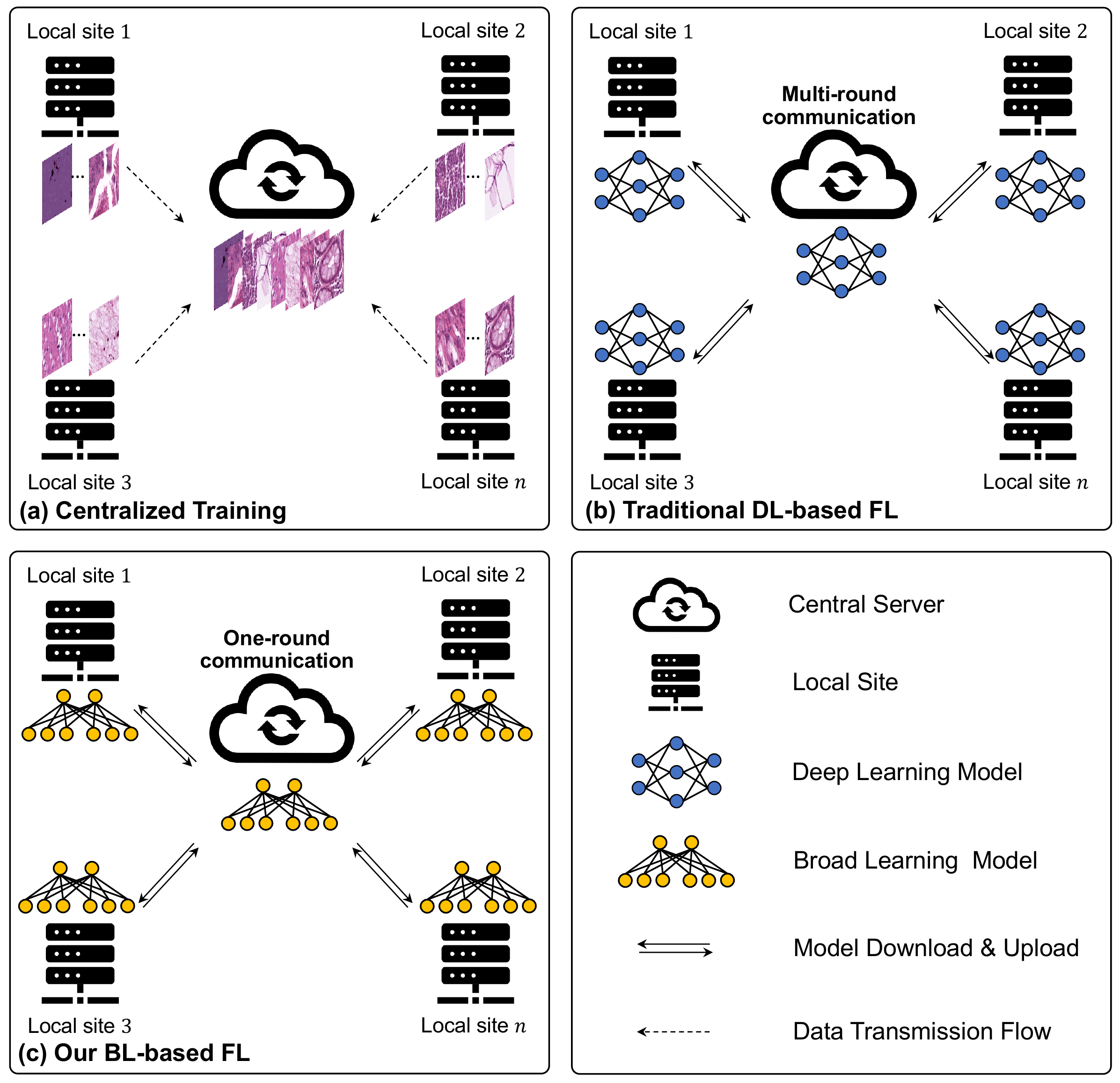}
	\caption{Overall comparison among centralized training, traditional DL-based FL and our proposed FedDBL paradigms. (a) Centralized learning gathers data from all the clients which cannot protect the patient's privacy. (b) Traditional DL-based FL preserves privacy by transmitting the model parameters to the central server without sharing the raw data. However, the communication overload highly depends on the model size and the number of communication rounds. (c) Our proposed FedDBL not only protects privacy, but also dramatically saves the communication burden through a super lightweight trainable broad learning system.}
	\label{fig:compare}
\end{figure}

Therefore, it is urged to construct a data-efficient and communication-efficient FL model for histopathological tissue classification. In this paper, we proposed a simple and effective solution for histopathological tissue classification, which considers not only the data sharing problem, but also the data dependency, communication efficiency, model robustness and model inversion attack. Our proposed model Federated Deep-Broad Learning, \textit{FedDBL} in short, contains three integrated components, including a common federated learning framework, a pre-trained deep learning (DL) backbone and a broad learning (BL) inference system~\cite{BLS,gong2022research}. The federated learning framework serves for decentralized training to avoid data sharing across different medical centers or institutions. The pre-trained DL backbone can provide stable and robust deep features when there are not enough training labels even with domain-irrelevant pre-trained DL backbone. It can also effectively avoid the model inversion attack since no back-propagation is calculated for gradients \cite{zhu2019deep}. The BL system is a lightweight classifier with good approximation capability which can greatly shorten the transmission time and overcome the data dependency problem. Fig.~\ref{fig:compare} comprehensively demonstrates the strengths of FedDBL compared with the centralized learning and the conventional federated learning ways.

Extensive experiments with five-fold cross-validation are conducted to demonstrate the superiority of FedDBL in several aspects, including data dependency, communication efficiency, flexibility and the practicability of the model encryption. When with enough training data, FedDBL can mostly outperform conventional FL strategies and achieve comparable or even better classification performance compared with centralized learning strategy. When reducing the training samples in the data dependency experiment, FedDBL still maintains a high-level performance and greatly outperforms both centralized learning and conventional FL frameworks, even with only 1\% training samples. FedDBL is also flexible to any deep learning architectures to support data- and communication-efficient histopathological tissue classification. Another spotlight of FedDBL is communication efficiency. Compared with the conventional FL frameworks, FedDBL's one-round training manner reduces the upload workload from 4.609GB to 276.5KB (over 17,000 times faster) with ResNet-50 backbone compared to traditional 50-round iterative training. Thanks to the tiny model size, FedDBL is also computationally efficient in model encryption which can further upgrade the privacy protection level. The main contributions of this paper can be summarized as follows:
\begin{itemize}
	\item We propose a novel federated learning approach (FedDBL) for histopathological tissue classification to preserve patients' privacy.
	\item To the best of our knowledge, FedDBL is the first study that considers communication efficiency and data efficiency simultaneously which reduces the communication overhead of each client by around 17,000 times on ResNet-50 with extremely limited training samples (only 1\%).
	\item FedDBL is a simple, effective and easy-to-use algorithm that associates three classical modules, including a robust pre-trained deep learning feature extractor, a fast broad learning inference system and a simple federated learning framework. It is highly extendable that allows to replace any module with a more advanced one.
	\item Extensive experiments demonstrate that FedDBL drastically relieves the dependence on training data and reduces the communication overhead while maintaining outstanding classification performance, which promotes its clinical practicability.
\end{itemize}

\section{Related Works}

\subsection{Histopathological Tissue Classification}
High-resolution WSIs offer a wide range of tissue phenotypes where the pixel-level annotation is time-consuming and requires a great deal of biomedical knowledge~(\cite{srinidhi2021deep}), making patch-level histopathological tissue classification an alternate solution for automated analysis in computer-aided tumor diagnosis~(\cite{kather2019predicting,xue2021selective,abdeltawab2021pyramidal}). 

Due to the rapid development of computer vision, the most popular natural image classification models can be transferred into histopathological tissue phenotyping. However, it still suffers from the data dependency problem with a huge annotation burden~(\cite{ayyad2021role}). Thus, various approaches have been proposed to reduce the annotation effort. \cite{han2022multi} proposed a multi-layer pseudo-supervision approach with a progressive dropout attention mechanism to convert patch-level labels into pseudo-pixel-level labels. An extra classification gate mechanism was presented which reduced the false-positive rate for non-predominant category classification and improved the segmentation performance in return. \cite{xue2021selective} utilized a generative adversarial network (GAN) to generate pseudo samples to expand the training data. \cite{dolezal2022uncertainty} cropped WSIs into tiles for training the uncertainty quantization model and solved the problem of domain shift in external validation data. In order to get rid of lacking image annotations, \cite{wang2022transformer} employed unsupervised contrastive learning to obtain a robust initialized model with moderate feature representation of the histopathological feature space, with no annotation burden. Our previous study (\cite{lin2022pdbl}) introduced pyramidal deep-broad learning (PDBL) as a pluggable module for any CNN backbone to further improve histopathological tissue classification performance.

Besides that, another unexplored challenge is the patient privacy issue. Only a few attempts~(\cite{saldanha2022swarm,saldanha2022direct}) have been made in federated learning for computational pathology, which will be discussed in the following subsection. And to the best of our knowledge, we are the first study to consider privacy protection in histopathological tissue classification.

\subsection{Federated Learning}
\subsubsection{Federated Learning in Medical Image Analysis}
Because of the ethical issue, federated learning (FL) has been widely adopted in medical applications to preserve the patients' privacy~(\cite{pati2022federated,warnat2021swarm,sheller2020federated}). In medical imaging, FL has witnessed a boost in interest (\cite{kaissis2020secure}), such as MRI reconstruction~(\cite{guo2021multi,li2020multi}), CT lesion segmentation~(\cite{yang2021federated}) and etc. 
In the COVID-19 pandemic, COVID-19-related applications with data from different medical centers or even from different countries become the most urgent demand in the real-world clinical scenario while FL greatly advances the diagnostic performance~(\cite{bai2021advancing}). \cite{dayan2021federated} used 20 institutes’ data across the global for predicting the future oxygen requirements of symptomatic patients suffering from COVID-19. \cite{dou2021federated} proposed a federated model to detect COVID-19 lung abnormalities with good generalization capability on unseen multinational datasets. 

\subsubsection{Federated Learning in Computational Histopathology}
In histopathological images, a swarm learning architecture with blockchain protocols has been proposed to predict the mutational status~(\cite{saldanha2022swarm}). However, compared with other medical imaging modalities, there are few studies~(\cite{saldanha2022direct}) that adopt federated learning in histopathological images for the following reasons. First, the digitalization of pathology is unpopular. Pathological diagnosis still relies on observing specimens under a microscope. Second, image annotation is also an obstacle for the histopathological image process since only pathologists are capable to label WSIs which greatly increases the difficulties of acquiring well-annotated data. Third, due to the gigapixel resolution of WSIs, the size of the deep learning model is generally large, which increases the communication burden in networking. 

There are technical solutions in FL to the high communication overhead problem, such as compressing the model size~(\cite{reisizadeh2020fedpaq,jhunjhunwala2021adaptive}). \cite{reisizadeh2020fedpaq} proposed FedPAQ to reduce the interactive overhead of FL by compressing the model with lower bit-precision and \cite{jhunjhunwala2021adaptive} proposed an adaptive quantization strategy to achieve communication efficiency. 

However, the underlying assumption of existing studies is that there should be enough samples for model training where they may not be able to take into account both communication efficiency and limited data issue~(\cite{kamp2021federated, zhang2023two}). In this study, we fully consider the specialty of histopathological images, the difficulties of data labeling and the communication efficiency in the real-world clinical scenario, which has never been discussed in decentralized computational pathology.
\section{Methodology}
In this section, we introduce our framework Federated Deep-Broad Learning (FedDBL). This framework is designed for privacy-preserving tissue classification with limited training samples and extremely low communication overhead. In the following subsections, we first describe the intuitive thinking and problem setting in Section~\ref{sub:problem-setting}. The overall framework and the methodology of FedDBL are shown in Section~\ref{sub:FedDBL}. Finally, we demonstrate the implementation details in Section~\ref{sub:implementation}.

\subsection{Problem Setting}
\label{sub:problem-setting}
As a classical upstream task in computational pathology, existing tissue classification approaches have achieved outstanding performance under an ideal condition with enough training samples by centralized learning. However, they might face the following obstacles in the real-world clinical scenario.

\textbf{Annotation burden:} Collecting enough well-labeled training samples is expensive and time-consuming because it requires labelers with medical background.  

\textbf{Privacy preservation:} The raw data should not be shared across different medical institutions (or clients) to preserve the patient's privacy. Transmitting raw data may break the principle of medical ethics.

\textbf{Communication cost:} The communication overhead has always been a challenge in federated learning models affected by many compound factors, such as the model size, the communication rounds, the model convergence speed, the network bandwidth and etc. 

\begin{figure*}[t]
	\centering
	\includegraphics[width=.975\linewidth]{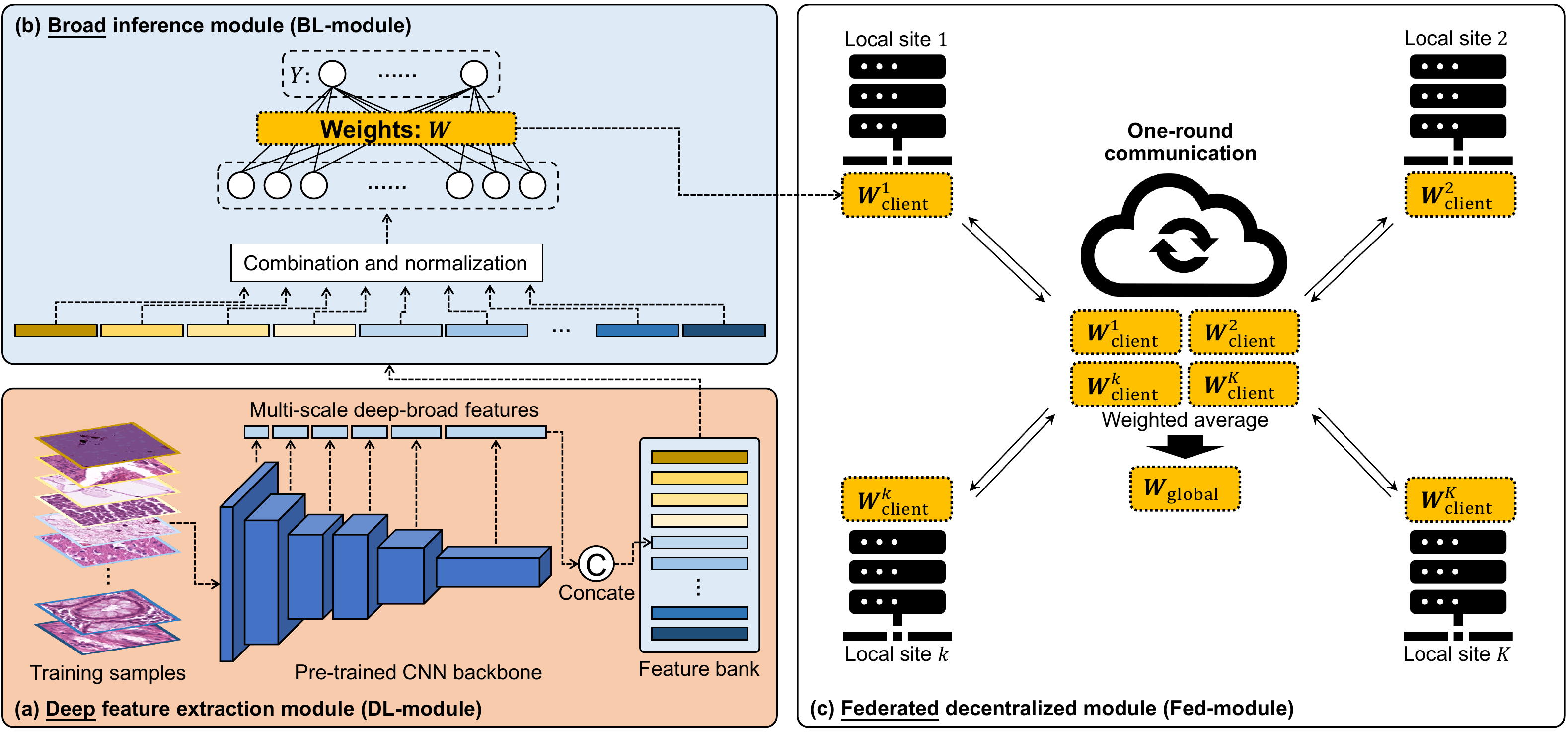}
	\caption{The overall architecture of FedDBL with three modules, deep feature extraction module, broad inference module and federated decentralized module. (a) Deep feature extraction module serves for extracting multi-scale deep-broad features from low level to high level by a pre-trained DL backbone. Features of all the patches are stored in a local feature bank. (b) Broad inference module is introduced for fast inference by a broad learning system. (c) Federated decentralized module applies a classical federated average approach to aggregate the broad learning weights from different clients.}
	\label{fig:FedDBL}
\end{figure*}

To resolve the aforementioned challenges, we propose a simple and effective FL-based framework, demonstrated in Fig.~\ref{fig:FedDBL}. First, we abandon conventional end-to-end training manner since limited training samples may harm the robustness of the deep learning model and decrease the convergence speed. Therefore, we separate feature extraction and inference for local training in each client. A pre-trained deep feature extractor (CNN backbone) is introduced to avoid the feature extractor being affected by the training sample bias from different clients in order to guarantee the robustness of extracted features. Then an independent broad learning inference system~\cite{BLS,lin2022pdbl} serves for fast inference. Finally, we apply a classical weighted averaging as in FedAvg~\cite{mcmahan2017communication}, to fuse the broad learning inference systems from all the clients.

\subsection{FedDBL Architecture and Formulation}
\label{sub:FedDBL}
As shown in Fig.~\ref{fig:FedDBL}, FedDBL consists of three modules, deep feature extraction module (DL-module), broad inference module (BL-module) and federated decentralized module (Fed-module). DL-module together with BL-module serves for local training on the client side. Fed-module is executed on the server side.
Algorithm~\ref{algorithm:FedDBL-server} provides the details of the entire FedDBL pipeline.

Let $\mathcal{D}_{1}, \mathcal{D}_{2}, \cdots, \mathcal{D}_k, \cdots, \mathcal{D}_{K}$ denote the local training sets from $K$ clients with the dataset size of $n_k$ for each client $\mathcal{D}_k$. The total number of training samples is denoted as $N = \sum_{k=1}^{K}{n_k}$. For each sample $X$ with ground truth $Y$ in $\mathcal{D}$, DL-module with pre-trained parameters $\Theta$ extracts the features and stores them in the local feature bank $\mathbf{B}$. Then BL-module calculates the weights $W_{client}$ of broad learning system. By the federated aggregation approach, we can obtain the global weight $W_{global}$. The workflows of the server and the clients are demonstrated in Algorithm~\ref{algorithm:FedDBL-server} and Algorithm~\ref{algorithm:FedDBL-client}, respectively.

\begin{algorithm}[!ht]
	\caption{FedDBL framework (Server Execution)}
	\label{algorithm:FedDBL-server}
	\SetKwInOut{Input}{Input}
	\SetKwInOut{Output}{Output}
	\Input  {A set of $K$ clients}
	\Output {A global model $W_{global}$}
	Prepare pre-trained DL backbone parameters $\Theta$
	
	Initialize BL system setting
	
	\For{each client $k$ \emph{\textbf{in parallel}}}{
		$W_{client}^{k} \leftarrow  $ClientExecution$\left(\Theta, \mathcal{D}_{k} \right)$
	}
	$W_{global} \leftarrow$ Fed-module$(W_{client}^1,\cdots,W_{client}^K)$
	
	\textbf{return} $W_{global}$
\end{algorithm}

\begin{algorithm}[!ht]
	\caption{FedDBL framework (Client Execution)}
	\label{algorithm:FedDBL-client}
	\SetKwInOut{Input}{Input}
	\SetKwInOut{Output}{Output}
	\Input  {Pre-trained DL backbone $\Theta$, training set $\mathcal{D}$ with $n$ training samples}
	\Output {Deep-broad learning model $W_{client}$.}
	
	\tcc{DL-module}
	\For{training sample $X$ in $\mathcal{D}$} {
		\For{$s$-th stage in $\Theta$}{
			$f_{X}^{s} \leftarrow \Theta^{s}\left(X\right)$ \tcp{Feature extraction}
			$\mathbf{e}_{X}^{s}=\frac{1}{H_{X}^{s} \times W_{X}^{s}} \sum_{i=1}^{H_{X}^{s}} \sum_{j=1}^{W_{X}^{s}} f_{X}^{s}(i,j)$ \tcp{Adaptive global average pooling}
			
		}
		$\mathbf{b}_{X} = \mathbf{e}_{X}^{1} \parallel \mathbf{e}_{X}^{2} \parallel \cdots \parallel \mathbf{e}_{X}^{S}$ \tcp{Concatenation}
		
	}
	\textbf{Obtain} $\left\{\mathbf{b}_{i}|i=1,2,\cdots,n \right\}$\\
	
	$\mathbf{B} \leftarrow \sigma\left( \left[\mathbf{b_1}^\mathbf{T}, \mathbf{b_2}^\mathbf{T}, \dots ,\mathbf{b}_{n}^\mathbf{T} \right]\right)^\mathbf{T}$ \tcp{Normalization transformation}
	
	\tcc{BL-module}
	Initialize BL system setting defined by central server \\
	
	$\mathbf{B}^{+} \leftarrow \lim_{\lambda \to 0}⁡ \left(\mathbf{B} \mathbf{B}^{\mathbf{T}} + \lambda E \right)^{-1} \mathbf{B}^{\mathbf{T}}$ \tcp{Solve Pseudo-inverse}
	
	$W_{client} \leftarrow \mathbf{B}^{+} Y$ \tcp{Calculate BL model weight}
	
	\textbf{return} $W_{client}$
	
\end{algorithm}

\subsubsection{Deep Feature Extraction Module}
A large number of samples and repeated backpropagation are required in standard DL training to achieve a good feature representation ability. When suffering from the insufficient data problem, the model training procedure might be unstable which leads to poor feature representation and model overfitting. Our previous study~\cite{lin2022pdbl} reveals that directly adopting a stable pre-trained model for feature extraction is more favorable to the model performance than training the model with limited samples, even the pre-trained model was trained by an irrelevant image domain (ImageNet\footnote{https://image-net.org/}). Inspired by this idea, we use a pre-trained CNN model with no further training to extract the deep features. Notice that, the selection of the pre-trained models is flexible, and can be from any image domain. We have conducted an experiment to justify the flexibility in Section~\ref{sec:exp}. Another advantage of using pre-trained models is to avoid model inverse attacks since the training samples are all unseen. To enrich the feature representation, we extract multi-stage features from low-level to high-level, details as below.

As illustrated in DL-module of Algorithm~\ref{algorithm:FedDBL-client}, each client $k$ $(k \in \left[1, \cdots, K\right])$ downloads the pre-trained DL backbone as feature extractor $\Theta$ and extracts multi-stage deep features $\mathbf{b}_X$ of training sample $X$ locally (we omit $k$ for simplicity), where $\mathbf{b}_X$ consists of multiple stages’ features $\Theta^s(X)(s \in \left[1, \cdots, S\right])$. The features of the entire dataset  $\mathcal{D}_k$ are stored in the local feature bank $\mathbf{B}$. Then the local feature bank $\mathbf{B}$ will be passed to broad inference module. Since neither the training data nor the feature bank is shared across different clients, there is no privacy leakage risk for the RAW data in deep feature extraction module. 

\subsubsection{Broad Inference Module}
With the local feature bank $\mathbf{B}$, each client $k$ can conduct a local BL system~\cite{BLS} through BL-module (Algorithm~\ref{algorithm:FedDBL-client}) for fast inference. By solving the Eq.~\eqref{eq:W_opt_1} optimization problem, an optimal BL model $W_{client}$ can be obtained rapidly through the pseudo-inverse method (Eq.~\eqref{eq:W_opt_2}).
\begin{equation}
\label{eq:W_opt_1}
W_{client}=\underset{W_{init}}{{\arg\min}} \left\| \mathbf{B}W_{init}-Y \right\|_{2}^{2} + \gamma \left\|W_{init} \right\|_{2}^{2}
\end{equation}
\begin{equation}
\label{eq:W_opt_2}
W_{client} = \mathbf{B}^{+} Y = \lim_{\lambda \to 0}⁡ \left(\mathbf{B} \mathbf{B}^{\mathbf{T}} + \lambda E \right)^{-1} \mathbf{B}^{\mathbf{T}} Y
\end{equation}
where $Y$ represents the ground-truth label matrix, $\mathbf{B}$ is local feature bank in the form of matrix. $W_{init}$ is the initialized broad learning weights. $E$ is the identity matrix, $\lambda$ is a constant parameter and $\gamma$ is the regularization parameter. The pseudo-inverse method of solving BL model considerably reduces the computational burden while achieving high communication efficiency. For the inference process, the predicted results can be calculated by $Y_{test}=\mathbf{B}_{test} W_{client}$ after extracting test samples’ deep features with the largest probabilistic value.

Thanks to the lightweight broad learning model $W_{client}$, the communication efficiency is drastically improved compared with the conventional DL-based FL frameworks.  

\subsubsection{Federated Decentralized Module}
In this module, we conduct a federated learning framework for decentralized learning. Given the broad learning model $W_{client}^k$ of each client $k$, we first upload the models from all the clients to the central server. And then general federated aggregation methods can be applied to aggregate them. Here, we use the most common weighted averaging way for model aggregation as adopted in FedAvg~\cite{mcmahan2017communication}, FedProx~\cite{li2020federated} and FedPAQ~\cite{reisizadeh2020fedpaq}.
\begin{equation}
\label{eq:W_global}
W_{global}=\sum_{k=1}^{K} \frac{n_{k}}{N}  W_{client}^{k}
\end{equation}
where $W_{global}$ is the global model from the server, $n_k $ is the number of training samples in client $k$ and $N$ is the total number of training samples. A larger training dataset will contribute more to the global model. Since we only share the broad learning model for once, the communication efficiency and the patient's privacy are guaranteed.

\subsection{Implementation Details}
\label{sub:implementation}

All of our experiments are implemented in Pytorch on a workstation with an NVIDIA RTX 3090 and the i9-11900K CPU with 16 cores. We use the cross-entropy loss for the baseline centralized training with batch size $20$. The SGD optimizer is set as follows: the learning rate is $1e^{-3}$, the weight decay is $1e^{-4}$ and the momentum is 0.9. The patches are $224 \times 224$ under $20\times$ WSIs. Different client numbers are used depending on the datasets.

We adopt three well-known federated aggregation methods, FedAvg \cite{mcmahan2017communication}, FedProx \cite{li2020federated} and FedPAQ \cite{reisizadeh2020fedpaq}, for comparison. And the centralized model is trained as the baseline. FedProx has the parameter $\mu$ to adjust the effect of the proximal term on the loss function. Here we set $\mu$ as $1$ which has better performance.
\section{Experiments}
\label{sec:exp}
In this section, we present the details of the datasets and conduct various experiments to demonstrate the performance and efficiency of the proposed \emph{FedDBL}. Section~\ref{sub:datasets} shows two open datasets and the experimental settings in the federated learning framework. In Section~\ref{sub:one-round}, we compare \emph{FedDBL} with centralized learning baselines, conventional federated learning baselines and one-round federated learning baselines. The effectiveness is comprehensively discussed in Section~\ref{sub:multi-round}. We use Matthews Correlation Coefficient (MCC), Accuracy and F1-score as the evaluation metrics in all the experiments.

\subsection{Datasets and Experimental Settings}
\label{sub:datasets}

\begin{table*}[ht]
	\centering
	\caption{Statistics of MC-CRC. \#1 denotes TCGA, \#2 denotes Kather, \#3 denotes Guangdong Provincial People’s Hospital and \#4 denotes Yunnan Cancer Hospital.}
	\begin{threeparttable}
		\begin{tabular}{p{0.04\textwidth}>{\makecell[r]}
				p{0.06\textwidth}>{\makecell[r]}
				p{0.06\textwidth}>{\makecell[r]}
				p{0.06\textwidth}>{\makecell[r]}
				p{0.06\textwidth}>{\makecell[r]}
				p{0.06\textwidth}>{\makecell[r]}
				p{0.06\textwidth}>{\makecell[r]}
				p{0.06\textwidth}>{\makecell[r]}
				p{0.06\textwidth}>{\makecell[r]}
				p{0.06\textwidth}>{\makecell[r]}
				p{0.07\textwidth}}
			\toprule
			&ADI &BACK &DEB &LYM &MUC &MUS &NORM &STR &TUM &\textbf{Total} \cr
			\midrule
			
			\#1\centering 	&\makecell[r]{10,065}	&\makecell[r]{10,736}	&\makecell[r]{10,603}	&\makecell[r]{2,340}	
			&\makecell[r]{9,398}	&\makecell[r]{12,974}	&\makecell[r]{10,003}	&\makecell[r]{10,081}	
			&\makecell[r]{12,899}	&\makecell[r]{89,099} \cr
			\#2\centering 	&\makecell[r]{10,407}	&\makecell[r]{10,566}	&\makecell[r]{11,512}	&\makecell[r]{11,577}	
			&\makecell[r]{8,896}	&\makecell[r]{13,536}	&\makecell[r]{8,763}	&\makecell[r]{10,446}	
			&\makecell[r]{14,317}	&\makecell[r]{100,000} \cr
			\#3\centering 	&\makecell[r]{10,000}	&\makecell[r]{22,565}	&\makecell[r]{9,999}	&\makecell[r]{5,831}	
			&\makecell[r]{10,737}	&\makecell[r]{10,000}	&\makecell[r]{13,368}	&\makecell[r]{12,584}	
			&\makecell[r]{10,000}	&\makecell[r]{105,084} \cr
			\#4\centering 	&\makecell[r]{2,500}	&\makecell[r]{2,500}	&\makecell[r]{2,500}	&\makecell[r]{2,500}	
			&\makecell[r]{2,500}	&\makecell[r]{2,500}	&\makecell[r]{2,500}	&\makecell[r]{2,500}	
			&\makecell[r]{2,500}	&\makecell[r]{22,500} \cr
			\midrule
			\textbf{Total}\centering &\makecell[r]{32,972}	&\makecell[r]{46,367}	&\makecell[r]{34,614}	&\makecell[r]{22,228}	
			&\makecell[r]{31,531}	&\makecell[r]{39,010}	&\makecell[r]{34,634}	&\makecell[r]{35,611}	
			&\makecell[r]{39,716}	&\makecell[r]{316,683} \cr
			\bottomrule
			\hspace{1mm}
		\end{tabular}
	\end{threeparttable}
	\label{tab:Multi-CRC}
\end{table*}
\begin{table}[t]
	\centering
	\caption{Statistics of BCSS. \#1, \#2 and \#3 are the datasets of three clients.}
	\begin{threeparttable}
		\begin{tabular}{p{0.03\textwidth}>{\makecell[r]}
				p{0.05\textwidth}>{\makecell[r]}
				p{0.05\textwidth}>{\makecell[r]}
				p{0.05\textwidth}>{\makecell[r]}
				p{0.05\textwidth}>{\makecell[r]}
				p{0.05\textwidth}}
			\toprule
			&TUM &STR &LYM &NEC &\textbf{Total} \cr
			\midrule
			
			\#1\centering &\makecell[r]{2,016}	&\makecell[r]{598}	&\makecell[r]{220}	&\makecell[r]{217}	&\makecell[r]{3,051} \cr
			\#2\centering &\makecell[r]{1,962}	&\makecell[r]{987}	&\makecell[r]{269}	&\makecell[r]{372}	&\makecell[r]{3,590} \cr
			\#3\centering &\makecell[r]{718}	&\makecell[r]{704}	&\makecell[r]{127}	&\makecell[r]{88}	&\makecell[r]{1,637} \cr
			\midrule
			\textbf{Total}\centering  &\makecell[r]{4,696}	&\makecell[r]{2,289}	&\makecell[r]{616}	&\makecell[r]{677}	&\makecell[r]{8,278}\cr
			\bottomrule
			\hspace{1mm}
		\end{tabular}
	\end{threeparttable}
	\label{tab:BCSS}
\end{table}

\textbf{Multi-center CRC (MC-CRC):} This is a multi-center datasets~\cite{zhao2020artificial,kather2019predicting} of colorectal cancer. Kather dataset \cite{kather2019predicting} defined nine different tissue types of H\&E stained WSIs, including adipose (ADI), background (BACK), debris (DEB), lymphocytes (LYM), mucus (MUC), smooth muscle (MUS), normal colon mucosa (NORM), cancer-associated stroma (STR), and colorectal adenocarcinoma epithelium (TUM). It contains 100k patches extracted from 86 WSIs. Following Kather, \cite{zhao2020artificial} also released another CRC dataset from three different medical centers, including 89.1k patches (85 slides) from The Cancer Genome Atlas (TCGA), 105.1k patches (106 slides) from Guangdong Provincial People's Hospital and 22.5k patches (48 slides) from Yunnan Cancer Hospital. All these patches are with the same resolution of $224\times 224$ at $20\times$ magnification. Table~\ref{tab:Multi-CRC} demonstrates the statistics of each dataset.

\textbf{BCSS}: Here, we introduce another dataset of breast cancer. Breast Cancer Semantic Segmentation (BCSS)~\cite{amgad2019structured} is an open challenge released in Grand Challenge\footnote{https://bcsegmentation.grand-challenge.org/}. There are 151 ROI images with pixel-level annotations in WSIs retrieved in TCGA. According to the naming convention provided by the supplementary document of BCSS, the ROIs are extracted from 21 different medical centers/hospitals. To generate a patch-level dataset, we first divide these ROIs into three clients and each of them has 7 medical centers. Then we crop each ROI into $224\times 224$ pixels by a sliding window with a step size of 120 pixels at $20\times$ objective magnification. Since this dataset is long-tail, we only keep the tissues from the four predominant classes, including tumor (TUM), stroma (STR), lymphocytic infiltrate (LYM), and necrosis or debris (NEC). The patches with the area of the majority class larger than 95\% are kept while the others are discarded as ambiguous patches. Finally, a total number of 8278 patches are left, and the size of each client’s dataset is shown in Table ~\ref{tab:BCSS}.

\textbf{Experimental Settings:}
We conduct the federated learning environment by the following steps. First of all, MC-CRC includes four clients according to the dataset setting from the original papers. BCSS is separated into three clients due to the limited training samples. For each client, the local dataset is randomly separated into a training set and a test set with a ratio of 7 : 3. To conduct a 5-fold cross-validation experiment, we repeat the dataset split for five times with different random seeds. Then, we randomly sample seven incremental subsets with the proportions of [1\%, 5\%, 10\%, 30\%, 50\%, 70\%, 100\%] from the training set for each dataset split. In addition, we combine the training sets as well as the test sets from all the clients for centralized learning comparison.

\begin{table*}[th]
	\centering
	\caption{Average MCC on MC-CRC dataset with different proportions of the training data under one-round training. We report the results of five-fold cross-validation with their 95\% confidence intervals. Centralized methods are used for baselines, \textbf{bold} is the highest among federated algorithms and \textcolor{red}{red} represents the highest among all methods including centralized learning. Prefix \emph{R-} and \emph{E-} indicate ResNet-50 and EfficientViT backbones pre-trained on ImageNet, respectively. Prefix \emph{C-} indicates CTransPath backbone pre-trained on pathology images.}
	\begin{threeparttable}
		\setlength{\tabcolsep}{2.5mm}{
			\begin{tabular}{p{0.09\textwidth}>{\centering}
					p{0.09\textwidth}>{\centering}
					p{0.09\textwidth}>{\centering}
					p{0.09\textwidth}>{\centering}
					p{0.09\textwidth}>{\centering}
					p{0.09\textwidth}>{\centering}
					p{0.09\textwidth}>{\centering}
					p{0.09\textwidth}}
				\toprule
				\multirow{2}{*}{Models} & \multicolumn{7}{c}{MCC under One-round Training on MC-CRC} \cr 
				\cmidrule(lr){2-8} 
				&1\% &5\% &10\% &30\% &50\% &70\% &100\%  \cr
				\midrule
				\multirow{2}{*}[1.2ex]{\emph{R-Centralized}} 	&0.6749 \\ (0.671, 0.678)
														&0.7718	\\ (0.771, 0.779)
														&0.8678	\\ (0.867, 0.871)
														&0.9060	\\ (0.901, 0.903)
														&0.9221	\\ (0.921, 0.923)
														&0.9344	\\ (0.933, 0.935)
														&0.9383	\\ (0.937, 0.939) \cr
				\multirow{2}{*}[1.2ex]{\emph{R-Centralized-FC}}	&0.8136	\\ (0.813, 0.815)
																&\textcolor{red}{0.8842} \\ (0.884, 0.885)
																&0.9019	\\ (0.902, 0.902)
																&0.9169	\\ (0.916, 0.917)
																&0.9299	\\ (0.929, 0.930)
																&0.9338	\\ (0.934, 0.934)
																&\textcolor{red}{0.9390} \\ (0.939, 0.939)	\cr
				\multirow{2}{*}[1.2ex]{\emph{R-FedAvg}} 	&0.0734	\\ (0.070, 0.075)
														&0.2499	\\ (0.243, 0.256)
														&0.3722	\\ (0.370, 0.379)
														&0.3812	\\ (0.377, 0.383)
														&0.4018	\\ (0.400, 0.403)
														&0.4336	\\ (0.433, 0.437)
														&0.3866 \\ (0.383, 0.387)	\cr
				\multirow{2}{*}[1.2ex]{\emph{R-FedProx}}     &0.1261	\\ (0.124, 0.134)
														&0.2627	\\ (0.259, 0.267)
														&0.4066	\\ (0.405, 0.410)
														&0.4203	\\ (0.420, 0.424)
														&0.4228	\\ (0.419, 0.425)
														&0.4013	\\ (0.401, 0.403)
														&0.4424 \\ (0.441, 0.447)	\cr
				\multirow{2}{*}[1.2ex]{\emph{R-FedPAQ}}		&0.1248	\\ (0.121, 0.127)
														&0.2875	\\ (0.286, 0.291)
														&0.3386	\\ (0.336, 0.340)
														&0.4806	\\ (0.479, 0.483)
														&0.4321	\\ (0.429, 0.433)
														&0.4473	\\ (0.446, 0.450)
														&0.4549 \\ (0.454, 0.460)	\cr
				\multirow{2}{*}[1.2ex]{\emph{R-FedAvg-FC}}   	&0.7709	\\ (0.770, 0.771)
														&\textbf{0.8375} \\ (0.837, 0.838)
														&0.8459	\\ (0.845, 0.846)
														&0.8672	\\ (0.866, 0.867)
														&0.8764	\\ (0.875, 0.876)
														&0.8720	\\ (0.872, 0.872)
														&0.8852 \\ (0.885, 0.886)	\cr
				\multirow{2}{*}[1.2ex]{\emph{R-FedDBL(ours)}} 	&\textbf{\textcolor{red}{0.8687}} \\ (0.868, 0.869)
																&0.8266	\\ (0.826, 0.827)
																&\textbf{\textcolor{red}{0.9136}} \\ (0.913, 0.914)
																&\textbf{\textcolor{red}{0.9341}} \\ (0.934, 0.934)
																&\textbf{\textcolor{red}{0.9340}} \\ (0.934, 0.934)
																&\textbf{\textcolor{red}{0.9342}} \\ (0.934, 0.934)
																&\textbf{0.9326} \\ (0.932, 0.933)	\cr
				
				\cdashline{1-8}
				\multirow{2}{*}[1.2ex]{\emph{E-Centralized}} 		&0.7772 \\ (0.776, 0.779)
															&0.8875 \\ (0.886, 0.889)
															&0.9092 \\ (0.907, 0.910)
															&\textcolor{red}{0.9401} \\ (0.940, 0.941)
															&\textcolor{red}{0.9434} \\ (0.941, 0.945)
															&\textcolor{red}{0.9535} \\ (0.953, 0.954)
															&\textcolor{red}{0.9565} \\ (0.956, 0.957)	\cr
				\multirow{2}{*}[1.2ex]{\emph{E-Centralized-FC}}    &0.7838 \\ (0.783, 0.784)
															&0.8671 \\ (0.867, 0.868)
															&0.8877 \\ (0.887, 0.888)
															&0.9146 \\ (0.914, 0.915)
															&0.9230 \\ (0.923, 0.923)
															&0.9261 \\ (0.926, 0.926)
															&0.9289 \\ (0.928, 0.929)	\cr
				\multirow{2}{*}[1.2ex]{\emph{E-FedAvg}}      	&0.6602 \\ (0.656, 0.665)
															&0.7669 \\ (0.765, 0.768)
															&0.7723 \\ (0.772, 0.775)
															&0.8008 \\ (0.800, 0.804)
															&0.7930 \\ (0.791, 0.794)
															&0.8001 \\ (0.799, 0.801)
															&0.8037 \\ (0.803, 0.804)	\cr
				\multirow{2}{*}[1.2ex]{\emph{E-FedProx}}     	&0.6918 \\ (0.690, 0.694)
														&0.7670 \\ (0.762, 0.768)
														&0.7499 \\ (0.747, 0.750)
														&0.7764 \\ (0.775, 0.779)
														&0.8010 \\ (0.800, 0.801)
														&0.7926 \\ (0.792, 0.795)
														&0.8027 \\ (0.801, 0.804)	\cr
				\multirow{2}{*}[1.2ex]{\emph{E-FedPAQ}}		&0.6506 \\ (0.647, 0.652)
															&0.7692 \\ (0.767, 0.772)
															&0.7602 \\ (0.759, 0.765)
															&0.7898 \\ (0.788, 0.790)
															&0.8180 \\ (0.817, 0.819)
															&0.8086 \\ (0.808, 0.812)
															&0.8006 \\ (0.799, 0.802)	\cr
				\multirow{2}{*}[1.2ex]{\emph{E-FedAvg-FC}}   	&0.6863 \\ (0.686, 0.688)
															&0.8079 \\ (0.807, 0.808)
															&0.8347 \\ (0.834, 0.835)
															&0.8671 \\ (0.867, 0.867)
															&0.8779 \\ (0.877, 0.878)
															&0.8846 \\ (0.884, 0.885)
															&0.8887 \\ (0.889, 0.889)	\cr
				\multirow{2}{*}[1.2ex]{\emph{E-FedDBL(ours)}}  	&\textcolor{red}{\textbf{0.8326}} \\ (0.832, 0.833)
															&\textcolor{red}{\textbf{0.9024}} \\ (0.902, 0.903)
															&\textcolor{red}{\textbf{0.9185}} \\ (0.918, 0.919)
															&\textbf{0.9247} \\ (0.924, 0.925)
															&\textbf{0.9244} \\ (0.924, 0.925)
															&\textbf{0.9243} \\ (0.924, 0.925)
															&\textbf{0.9246} \\ (0.924, 0.925)	\cr

				\cdashline{1-8}
				\multirow{2}{*}[1.2ex]{\emph{C-Centralized-FC}}		&\textcolor{red}{0.9314} \\ (0.931, 0.932)
															&0.9542	\\ (0.954, 0.954)
															&\textcolor{red}{0.9628} \\ (0.962, 0.963)
															&\textcolor{red}{0.9724} \\ (0.972, 0.973)
															&\textcolor{red}{0.9761} \\ (0.976, 0.976)
															&\textcolor{red}{0.9775} \\ (0.977, 0.978)
															&\textcolor{red}{0.9793} \\ (0.979, 0.980)	\cr
				\multirow{2}{*}[1.2ex]{\emph{C-FedAvg-FC}} 		&0.8964	\\ (0.896, 0.897)
														&0.9306 \\ (0.930, 0.931)
														&0.9387	\\ (0.938, 0.939)
														&0.9478	\\ (0.947, 0.948)
														&0.9507	\\ (0.950, 0.951)
														&0.9524	\\ (0.952, 0.953)
														&0.9543	\\ (0.954, 0.955)	\cr
				\multirow{2}{*}[1.2ex]{\emph{C-FedDBL(ours)}} &\textbf{0.9114} \\ (0.911, 0.912)
														&\textbf{\textcolor{red}{0.9594}} \\ (0.959, 0.960)
														&\textbf{0.9610} \\ (0.961, 0.962)
														&\textbf{0.9620} \\ (0.962, 0.963)
														&\textbf{0.9625} \\ (0.962, 0.963)
														&\textbf{0.9627} \\ (0.962, 0.963)
														&\textbf{0.9627} \\ (0.962, 0.963)	\cr
				\bottomrule
				\hspace{1mm}
		\end{tabular}}
	\end{threeparttable}
	\label{tab:CRC_one-round-mcc}
\end{table*}

\begin{table*}[th]
	\centering
	\caption{Average MCC on BCSS dataset with different proportions of the training data under one-round training. For simplification, the detail settings can be found in Table~\ref{tab:CRC_one-round-mcc}.}
	\begin{threeparttable}
		\setlength{\tabcolsep}{2.5mm}{
			\begin{tabular}{p{0.09\textwidth}>{\centering}
					p{0.09\textwidth}>{\centering}
					p{0.09\textwidth}>{\centering}
					p{0.09\textwidth}>{\centering}
					p{0.09\textwidth}>{\centering}
					p{0.09\textwidth}>{\centering}
					p{0.09\textwidth}>{\centering}
					p{0.09\textwidth}}
				\toprule
				\multirow{2}{*}{Models} & \multicolumn{7}{c}{MCC under One-round Training on BCSS} \cr 
				\cmidrule(lr){2-8} 
				&1\% &5\% &10\% &30\% &50\% &70\% &100\%  \cr
				\midrule
				\multirow{2}{*}[1.2ex]{\emph{R-Centralized}}	&0.6043 \\ (0.592, 0.608)
														&0.1734 \\ (0.167, 0.175)
														&0.4162 \\ (0.406, 0.432)
														&0.7998 \\ (0.799, 0.817)
														&0.8185 \\ (0.817, 0.822)	
														&0.8403 \\ (0.835, 0.843)
														&0.8964 \\ (0.895, 0.897)	\cr
				\multirow{2}{*}[1.2ex]{\emph{R-Centralized-FC}}	&0.0903	\\ (0.085, 0.099)
															&0.6661	\\ (0.666, 0.668)
															&0.7646	\\ (0.763, 0.766)
															&0.8880	\\ (0.887, 0.888)
															&0.9134	\\ (0.913, 0.914)
															&0.9223	\\ (0.922, 0.923)
															&0.9318 \\ (0.931, 0.932)	\cr
				\multirow{2}{*}[1.2ex]{\emph{R-FedAvg}}	&0.1728 \\ (0.160, 0.176)
													&0.5569 \\ (0.555, 0.561)
													&0.4704 \\ (0.470, 0.483)
													&0.4328 \\ (0.417, 0.438)
													&0.3122 \\ (0.305, 0.335)
													&0.1456 \\ (0.137, 0.151)
													&0.2998 \\ (0.291, 0.315)	\cr
				\multirow{2}{*}[1.2ex]{\emph{R-FedProx}}	&0.1602 \\ (0.147, 0.164)
													&0.5153 \\ (0.510, 0.520)
													&0.4453 \\ (0.427, 0.456)
													&0.4771 \\ (0.464, 0.483)
													&0.3311 \\ (0.323, 0.352)	
													&0.3621 \\ (0.345, 0.365)
													&0.2307 \\ (0.208, 0.236)	\cr
				\multirow{2}{*}[1.2ex]{\emph{R-FedPAQ}}	&0.1453	\\ (0.131, 0.147)
													&0.3840	\\ (0.363, 0.394)
													&0.3004	\\ (0.278, 0.310)
													&0.4132	\\ (0.412, 0.420)
													&0.2364	\\ (0.232, 0.239)
													&0.3804	\\ (0.361, 0.387)
													&0.2005 \\ (0.197, 0.213)	\cr
				\multirow{2}{*}[1.2ex]{\emph{R-FedAvg-FC}}	&0.0687 \\ (0.064, 0.072)
													&0.2514 \\ (0.236, 0.255)
													&0.5959 \\ (0.593, 0.598)
													&0.7848 \\ (0.784, 0.786)
													&0.8309 \\ (0.830, 0.832)
													&0.8731 \\ (0.873, 0.874)
													&0.8915 \\ (0.891, 0.892)	\cr
				\multirow{2}{*}[1.2ex]{\emph{R-FedDBL(ours)}}	&\textbf{\textcolor{red}{0.8314}} \\ (0.829, 0.832)
															&\textbf{\textcolor{red}{0.9165}} \\ (0.916, 0.917)
															&\textbf{\textcolor{red}{0.9323}} \\ (0.931, 0.932)
															&\textbf{\textcolor{red}{0.9508}} \\ (0.950, 0.951)
															&\textbf{\textcolor{red}{0.9566}} \\ (0.956, 0.957)
															&\textbf{\textcolor{red}{0.9542}} \\ (0.954, 0.954)
															&\textbf{\textcolor{red}{0.9408}} \\ (0.940, 0.941)	\cr
				
				\cdashline{1-8}
				\multirow{2}{*}[1.2ex]{\emph{E-Centralized}} 	&0.3459 \\ (0.343, 0.349)
																&0.7647 \\ (0.759, 0.768)
																&0.8523 \\ (0.849, 0.854)
																&0.8052 \\ (0.801, 0.810)
																&0.8833 \\ (0.879, 0.884)
																&0.8877 \\ (0.883, 0.888)
																&0.9090	\\ (0.906, 0.911)	\cr
				\multirow{2}{*}[1.2ex]{\emph{E-Centralized-FC}}      	&0.2711 \\ (0.263, 0.272)
																		&0.2686 \\ (0.263, 0.270)
																		&0.6224 \\ (0.621, 0.626)
																		&0.7885 \\ (0.787, 0.789)
																		&0.8501 \\ (0.849, 0.851)
																		&0.8874 \\ (0.887, 0.888)
																		&0.9072 \\ (0.906, 0.907)	\cr
				\multirow{2}{*}[1.2ex]{\emph{E-FedAvg}}      	&0.0205 \\ (0.018, 0.027)
																&0.4393 \\ (0.433, 0.446)
																&0.6270 \\ (0.623, 0.637)
																&0.8947 \\ (0.893, 0.895)
																&0.8867 \\ (0.884, 0.887)
																&0.9060 \\ (0.905, 0.908)
																&0.8946 \\ (0.894, 0.897)	\cr
				\multirow{2}{*}[1.2ex]{\emph{E-FedProx}}     	&0.0356 \\ (0.034, 0.042)
																&0.4540 \\ (0.447, 0.459)
																&0.6828 \\ (0.682, 0.690)
																&0.8912 \\ (0.889, 0.891)
																&0.8869 \\ (0.885, 0.889)
																&0.9052 \\ (0.903, 0.906)
																&0.8896	\\ (0.889, 0.892)	\cr
				\multirow{2}{*}[1.2ex]{\emph{E-FedPAQ}}			&0.0883 \\ (0.082, 0.089)
																&0.4511 \\ (0.449, 0.458)
																&0.6981 \\ (0.689, 0.700)
																&0.8990 \\ (0.898, 0.900)
																&0.8979 \\ (0.897, 0.899)
																&0.9052 \\ (0.904, 0.906)
																&0.8932 \\ (0.892, 0.895)	\cr
				\multirow{2}{*}[1.2ex]{\emph{E-FedAvg-FC}}   	&0.1453 \\ (0.135, 0.150)
																&0.1558 \\ (0.148, 0.166)
																&0.0911 \\ (0.088, 0.092)
																&0.6292 \\ (0.627, 0.630)
																&0.6897 \\ (0.689, 0.690)
																&0.7574 \\ (0.757, 0.758)
																&0.8045 \\ (0.804, 0.805)	\cr
				\multirow{2}{*}[1.2ex]{\emph{E-FedDBL(ours)}}  	&\textbf{\textcolor{red}{0.8241}} \\ (0.824, 0.825)
																	&\textbf{\textcolor{red}{0.8920}} \\ (0.891, 0.892)
																	&\textbf{\textcolor{red}{0.8846}} \\ (0.884, 0.885)
																	&\textbf{\textcolor{red}{0.9121}} \\ (0.911, 0.912)
																	&\textbf{\textcolor{red}{0.9330}} \\ (0.932, 0.933)
																	&\textbf{\textcolor{red}{0.9386}} \\ (0.938, 0.939)
																	&\textbf{\textcolor{red}{0.9414}} \\ (0.941, 0.941)	\cr

				\cdashline{1-8}
				\multirow{2}{*}[1.2ex]{\emph{C-Centralized-FC}}	&0.2409	\\ (0.234, 0.249)
															&0.1544	\\ (0.150, 0.163)
															&0.3983	\\ (0.395, 0.400)
															&0.8044	\\ (0.804, 0.805)
															&0.9033	\\ (0.903, 0.904)
															&0.9332	\\ (0.932, 0.933)
															&0.9533 \\ (0.952, 0.953)	\cr
				\multirow{2}{*}[1.2ex]{\emph{C-FedAvg-FC}} 		&0.1444	\\ (0.140, 0.149)
															&0.0871	\\ (0.084, 0.094)
															&0.0163	\\ (0.015, 0.017)
															&0.3982	\\ (0.396, 0.399)
															&0.6627	\\ (0.662, 0.663)
															&0.7461	\\ (0.746, 0.747)
															&0.8409 \\ (0.841, 0.842)	\cr
				\multirow{2}{*}[1.2ex]{\emph{C-FedDBL(ours)}}	&\textbf{\textcolor{red}{0.9308}} \\ (0.930, 0.931)
															&\textbf{\textcolor{red}{0.9582}} \\ (0.958, 0.959)
															&\textbf{\textcolor{red}{0.9621}} \\ (0.961, 0.962)
															&\textbf{\textcolor{red}{0.9634}} \\ (0.963, 0.963)
															&\textbf{\textcolor{red}{0.9718}} \\ (0.971, 0.972)
															&\textbf{\textcolor{red}{0.9802}} \\ (0.980, 0.981)
															&\textbf{\textcolor{red}{0.9847}} \\ (0.984, 0.985)	\cr
				
				\bottomrule
				\hspace{1mm}
		\end{tabular}}
	\end{threeparttable}
	\label{tab:BCSS_one-round-mcc}
\end{table*}

\subsection{Comparisons under One-round Communication}
\label{sub:one-round}
In this experiment, we evaluate the data efficiency, communication efficiency and flexibility of our proposed \emph{FedDBL}. Table~\ref{tab:CRC_one-round-mcc} and \ref{tab:BCSS_one-round-mcc} demonstrate the average MCC performance on MC-CRC and BCSS respectively, the results of average accuracy and F1-score can be found in Table~\ref{tab:CRC_one-round-accuracy}, Table~\ref{tab:BCSS_one-round-accuracy}, Table~\ref{tab:CRC_one-round-f1-score}, and Table~\ref{tab:BCSS_one-round-f1-score}. We compare \emph{FedDBL} with four FL frameworks and two centralized training approaches with only one-round communication or local training. In this experiment, we employ ResNet-50 (\textit{R-}) and EfficientViT-M4 (\textit{E-})\cite{liu2023efficientvit} pre-trained on ImageNet as domain-irrelevant DL-module, and CTransPath (\textit{C-}) pre-trained on pathology images as domain-specific DL-module for \emph{FedDBL} and all the other competitors. 
\begin{enumerate}
	\item \emph{Centralized}: We fine-tune the pre-trained backbone with a random initialized fully connected (FC) layer in the centralized learning manner.
	\item \emph{Centralized-FC}: We freeze the pre-trained backbone while fine-tuning the FC-layer in the centralized learning manner.
	\item \emph{FedAvg}: We fine-tune the pre-trained model by FedAvg framework~\cite{mcmahan2017communication}.
	\item \emph{FedProx}: We fine-tune the pre-trained model by FedProx framework~\cite{li2020federated}.
	\item \emph{FedPAQ}: We fine-tune the pre-trained model by a communication-efficient federated learning framework, FedPAQ~\cite{reisizadeh2020fedpaq}.
	\item \emph{FedAvg-FC}: We freeze the pre-trained CNN backbone and only update the FC-layer by FedAvg framework.
\end{enumerate}

\textbf{Data efficiency:} Let us take the experiment with ResNet-50 architecture (\emph{R-}) as the example. As shown in Table~\ref{tab:CRC_one-round-mcc} and \ref{tab:BCSS_one-round-mcc}, when with enough training samples (100\%) in one-round training experiment, centralized learning can achieve better performance in both datasets than other conventional FL frameworks, \emph{R-FedAvg}, \emph{R-FedProx} and \emph{R-FedPAQ}. Because centralized learning gathers the training samples from all clients such that the CNN model is trained more stably with a faster convergence speed than existing FL frameworks. When freezing the CNN backbone, \emph{R-FedAvg-FC} returns to a more favorable performance. \emph{R-Centralized-FC} also show better performance than \emph{R-Centralized}. This observation shows that when with limited communication resources or local training time but with sufficient training samples, maintaining a more stable CNN feature extractor is better than retraining the entire model. Only updating FC-layer is a better solution under this circumstance. The proposed \emph{R-FedDBL} can achieve comparable performance with centralized learning strategies in MC-CRC dataset and even outperform them in BCSS dataset. When reducing the training data, the performance of all the approaches drops dramatically except \emph{R-FedDBL}, especially when with only 1\% training samples. \emph{R-FedAvg-FC} with the frozen CNN backbone achieves around 0.77 MCC (0.79 accuracy and F1-score) in MC-CRC but is still less effective than \emph{R-FedDBL}. Similar conclusions can be observed when replacing the deep learning architecture with EfficientViT (\textit{E-}). \emph{E-FedDBL} shows more stable performance in extremely limited training samples compared to \emph{E-FedAvg-FC}. Moreover, the quantitative results of either \emph{R-FedAvg-FC} or \emph{E-FedAvg-FC} in BCSS with 1\% training samples are much worse than the ones in MC-CRC dataset because the amount of training samples in MC-CRC is around 38 times larger than BCSS. In this experiment, the proposed \emph{FedDBL} performs the most stable quantitative results among all the approaches in both datasets. It even outperforms centralized learning in most of the training data proportions. From this experiment, we can conclude that when with limited network communication resources and training samples, \emph{FedDBL} is the best solution for histopathological tissue classification.

\textbf{Flexibility:} Besides the data and communication efficiency, \emph{FedDBL} is also a flexible framework that can be further upgraded by replacing any module with a superior one if it exists, for example, a more robust feature extractor, a more outstanding classifier or a superior federated aggregation strategy. In this experiment, we demonstrate the flexibility of \emph{FedDBL} by replacing the CNN backbone pre-trained on ImageNet with a domain-specific backbone CTransPath~\cite{wang2022transformer} pre-trained on histopathological images. The lower parts of both datasets in Table~\ref{tab:CRC_one-round-mcc} and \ref{tab:BCSS_one-round-mcc} demonstrate the comparisons under CTransPath. Here, we only compare \emph{FedDBL} with the ones only updating FC-layer. When experimenting on the larger dataset of MC-CRC, the domain-specific pre-trained feature extractor CTransPath can greatly improve all three approaches. Centralized learning demonstrates the best results in almost all the dataset proportions. But \emph{C-FedDBL} still constantly outperforms \emph{C-FedAvg-FC} under the same circumstance. In the much smaller dataset of BCSS, \emph{C-FedDBL} demonstrates its superiority and outperforms both \emph{C-Centralized-FC} and \emph{C-FedAvg-FC}. When with only 1\% training samples in BCSS, CTransPath can improve the MCC of \emph{FedDBL} from 0.8314 (ResNet-50) to 0.9308. Less or no improvement is observed for the other two approaches.

\textbf{Communication efficiency:} Higher communication efficiency benefits not only from fewer communication rounds but also from a smaller model or feature size for transmission. Conventional federated frameworks share either the parameters of the deep learning models or the extracted features. In our proposed \emph{FedDBL}, we only share the lightweight BL-module weights without sharing any deep learning parameters or deep features. In Table~\ref{tab:Overhead}, we demonstrate the model size and total upload overhead per client of the entire training phase. We can observe that the size of ResNet-50 for sharing is 94.4MB for each communication round. \emph{R-FedDBL}, \emph{E-FedDBL} and \emph{C-FedDBL} share only 276.5KB, 55.4KB and 55.4KB broad learning weights respectively. With only one-round communication, \emph{R-FedDBL} reduces the communication overhead by nearly 350 times compared with vanilla ResNet-50. Since conventional federated frameworks might need multiple training iterations for model convergence, 50-round communication will greatly increase the total upload overhead from 94.4MB to 4.609GB. Thanks to the lightweight BL-module and one-shot training manner, \emph{E-FedDBL} and \emph{C-FedDBL} reduces the upload workload from 4.609GB to 55.4KB which is over 87,000 times faster than \emph{R-FedAvg} and over 31,600 times faster than \emph{E-FedAvg}. Even \emph{R-FedDBL} is over 17,000 times faster.

\begin{table*}[th]
	\centering
	\caption{Comparisons with different methods on 50-round training (MCC). For simplification, the detail settings can be found in Table~\ref{tab:CRC_one-round-mcc}. $\dagger$ means models trained for 50 rounds.}
	\begin{threeparttable}
		\setlength{\tabcolsep}{2.5mm}{
			\begin{tabular}{p{0.04\textwidth}>{}
					p{0.09\textwidth}>{\centering}
					p{0.09\textwidth}>{\centering}
					p{0.09\textwidth}>{\centering}
					p{0.09\textwidth}>{\centering}
					p{0.09\textwidth}>{\centering}
					p{0.09\textwidth}>{\centering}
					p{0.09\textwidth}>{\centering}
					p{0.09\textwidth}}
				\toprule
				\multirow{2}{*}{Datasets} & \multirow{2}{*}{Models} & \multicolumn{7}{c}{MCC under Multiple-round Training} \cr 
				\cmidrule(lr){3-9} 
				&	&1\% &5\% &10\% &30\% &50\% &70\% &100\%  \cr
				\midrule
				\multirow{18}{*}{\makecell[l]{MC-\\CRC}}
					&\multirow{2}{*}[1.2ex]{\emph{R-Centralized$\dagger$}} 	&0.8687	\\ (0.868, 0.869)
																		&0.9222	\\ (0.922, 0.923)
																		&0.9446	\\ (0.944, 0.945)
																		&0.9676 \\ (0.967, 0.968)
																		&0.9762 \\ (0.976, 0.977)
																		&0.9793 \\ (0.979, 0.980)
																		&0.9840 \\ (0.984, 0.984)	\cr
					&\multirow{2}{*}[1.2ex]{\emph{E-Centralized$\dagger$}} 	&0.8836 \\ (0.884, 0.886)
																		&0.9349 \\ (0.934, 0.935)
																		&0.9527 \\ (0.952, 0.953)
																		&\textcolor{red}{0.9698} \\ (0.969, 0.970)
																		&\textcolor{red}{0.9768} \\ (0.976, 0.977)
																		&\textcolor{red}{0.9817} \\ (0.981, 0.982)
																		&\textcolor{red}{0.9842} \\ (0.984, 0.984)	\cr

					\cdashline{2-9}
					&\multirow{2}{*}[1.2ex]{\emph{R-FedAvg$\dagger$}}		&0.8593	\\ (0.859, 0.860)
																	&0.9199	\\ (0.919, 0.920)
																	&0.9308	\\ (0.930, 0.931)
																	&0.9484	\\ (0.948, 0.949)
																	&0.9537	\\ (0.953, 0.954)
																	&0.9553	\\ (0.955, 0.955)
																	&0.9595 \\ (0.959, 0.960)	\cr
					&\multirow{2}{*}[1.2ex]{\emph{R-FedProx$\dagger$}}     &0.8637	\\ (0.863, 0.864)
																	&0.9199	\\ (0.919, 0.920)
																	&0.9335	\\ (0.933, 0.934)
																	&0.9491	\\ (0.949, 0.950)
																	&0.9533	\\ (0.953, 0.954)
																	&0.9504	\\ (0.950, 0.952)
																	&0.9574 \\ (0.957, 0.958)	\cr
					
					&\multirow{2}{*}[1.2ex]{\emph{E-FedAvg$\dagger$}}      &0.9070 \\ (0.907, 0.908)
																	&0.9361 \\ (0.936, 0.936)
																	&0.9428 \\ (0.942, 0.943)
																	&0.9572 \\ (0.957, 0.958)
																	&0.9610 \\ (0.961, 0.962)
																	&0.9217 \\ (0.916, 0.926)
																	&0.9663 \\ (0.966, 0.967)	\cr
					&\multirow{2}{*}[1.2ex]{\emph{E-FedProx$\dagger$}}		&0.9039 \\ (0.904, 0.905)
																	&0.9362 \\ (0.936, 0.936)
																	&0.9436 \\ (0.943, 0.944)
																	&0.9567 \\ (0.956, 0.957)
																	&0.9623 \\ (0.962, 0.963)
																	&\textbf{0.9653} \\ (0.965, 0.966)
																	&\textbf{0.9673} \\ (0.967, 0.968)	\cr
				
					&\multirow{2}{*}[1.2ex]{\emph{R-FedDBL(ours)}}	&0.8687	\\ (0.868, 0.869)
														&0.8266	\\ (0.826, 0.827)
														&0.9136	\\ (0.913, 0.914)
														&0.9341	\\ (0.934, 0.934)
														&0.9340	\\ (0.934, 0.934)
														&0.9342	\\ (0.934, 0.934)
														&0.9326 \\ (0.932, 0.933)	\cr
					&\multirow{2}{*}[1.2ex]{\emph{E-FedDBL(ours)}}	&0.8326 \\ (0.832, 0.833)
														&0.9024 \\ (0.902, 0.903)
														&0.9185 \\ (0.918, 0.919)
														&0.9247 \\ (0.924, 0.925)
														&0.9244 \\ (0.924, 0.925)
														&0.9243 \\ (0.924, 0.925)
														&0.9246 \\ (0.924, 0.925)	\cr
					&\multirow{2}{*}[1.2ex]{\emph{C-FedDBL(ours)}}  	&\textbf{\textcolor{red}{0.9114}} \\ (0.911, 0.912)
														&\textbf{\textcolor{red}{0.9594}} \\ (0.959, 0.960)
														&\textbf{\textcolor{red}{0.9610}} \\ (0.961, 0.962)
														&\textbf{0.9620} \\ (0.962, 0.963)
														&\textbf{0.9625} \\ (0.962, 0.963)
														&0.9627 \\ (0.962, 0.963)
														&0.9627 \\ (0.962, 0.963)		\cr
				
				\midrule
				\multirow{18}{*}{\centering BCSS}
					&\multirow{2}{*}[1.2ex]{\emph{R-Centralized$\dagger$}}		&0.7358	\\ (0.728, 0.740)
																		&0.8660	\\ (0.863, 0.867)
																		&0.9093	\\ (0.908, 0.911)
																		&0.9409	\\ (0.941, 0.943)
																		&0.9678	\\ (0.968, 0.968)
																		&0.9693	\\ (0.969, 0.970)
																		&0.9729 \\ (0.972, 0.974)	\cr
					&\multirow{2}{*}[1.2ex]{\emph{E-Centralized$\dagger$}} 	&0.7644 \\ (0.758, 0.765)
																		&0.9190 \\ (0.918, 0.920)
																		&0.9219 \\ (0.920, 0.923)
																		&0.9385 \\ (0.936, 0.940)
																		&0.9663 \\ (0.966, 0.967)
																		&0.9712 \\ (0.971, 0.971)
																		&0.9708 \\ (0.970, 0.972)	\cr
					
					\cdashline{2-9}
					&\multirow{2}{*}[1.2ex]{\emph{R-FedAvg$\dagger$}}		&0.7741	\\ (0.768, 0.779)
																	&0.8893	\\ (0.887, 0.890)
																	&0.8992	\\ (0.897, 0.901)
																	&0.9383	\\ (0.937, 0.939)
																	&0.9543	\\ (0.954, 0.955)
																	&0.9668	\\ (0.967, 0.967)
																	&0.9719 \\ (0.971, 0.972)	\cr
					&\multirow{2}{*}[1.2ex]{\emph{R-FedProx$\dagger$}}		&0.7912	\\ (0.788, 0.795)
																	&0.8710	\\ (0.868, 0.872)
																	&0.9171	\\ (0.914, 0.918)
																	&0.9445	\\ (0.944, 0.945)
																	&0.9557	\\ (0.955, 0.957)
																	&0.9620	\\ (0.962, 0.962)
																	&0.9712 \\ (0.971, 0.971)	\cr
					
					&\multirow{2}{*}[1.2ex]{\emph{E-FedAvg$\dagger$}}     	&0.7612 \\ (0.758, 0.763)
																	&0.9001 \\ (0.899, 0.901)
																	&0.9069 \\ (0.904, 0.908)
																	&0.9629 \\ (0.962, 0.963)
																	&0.9686 \\ (0.968, 0.969)
																	&0.9741 \\ (0.974, 0.975)
																	&0.9728 \\ (0.972, 0.973)	\cr
					&\multirow{2}{*}[1.2ex]{\emph{E-FedProx$\dagger$}}     &0.8058 \\ (0.800, 0.808)
																	&0.8971 \\ (0.895, 0.898)
																	&0.9280 \\ (0.926, 0.928)
																	&0.9598 \\ (0.959, 0.960)
																	&0.9634 \\ (0.963, 0.964)
																	&0.9744 \\ (0.974, 0.975)
																	&0.9749 \\ (0.974, 0.975)	\cr
					
					&\multirow{2}{*}[1.2ex]{\emph{R-FedDBL(ours)}}		&0.8314	\\ (0.829, 0.832)
															&0.9165	\\ (0.916, 0.917)
															&0.9323	\\ (0.931, 0.932)
															&0.9508	\\ (0.950, 0.951)
															&0.9566	\\ (0.956, 0.957)
															&0.9542	\\ (0.954, 0.954)
															&0.9408	\\ (0.940, 0.941)	\cr
					&\multirow{2}{*}[1.2ex]{\emph{E-FedDBL(ours)}}      	&0.8241 \\ (0.824, 0.825)
															&0.8920 \\ (0.891, 0.892)
															&0.8846 \\ (0.884, 0.885)
															&0.9121 \\ (0.911, 0.912)
															&0.9330 \\ (0.932, 0.933)
															&0.9386 \\ (0.938, 0.939)
															&0.9414 \\ (0.941, 0.941)	\cr
					&\multirow{2}{*}[1.2ex]{\emph{C-FedDBL(ours)}}		&\textbf{\textcolor{red}{0.9308}} \\ (0.930, 0.931)
															&\textbf{\textcolor{red}{0.9582}} \\ (0.958, 0.959)
															&\textbf{\textcolor{red}{0.9621}} \\ (0.961, 0.962)
															&\textbf{\textcolor{red}{0.9634}} \\ (0.963, 0.963)
															&\textbf{\textcolor{red}{0.9718}} \\ (0.971, 0.972)
															&\textbf{\textcolor{red}{0.9802}} \\ (0.980, 0.981)
															&\textbf{\textcolor{red}{0.9847}} \\ (0.984, 0.985)	\cr
				\bottomrule
				\hspace{1mm}
		\end{tabular}}
	\end{threeparttable}
	\label{tab:mcc}
\end{table*}

\begin{table}[t]
	\centering
	\caption{Model size and total upload overhead per client among three models.}
	\begin{threeparttable}
		\begin{tabular*}{0.35\textwidth}{@{\extracolsep{\fill}}lrr}
			\toprule
			Models\centering 	& Model size & \makecell[c]{Total upload size} \cr
			\midrule
			\emph{R-FedAvg} 	&94.4MB 	&4.609GB \cr
			\emph{E-FedAvg} 	&34.2MB		&1.670GB \cr
			\emph{R-FedDBL} 	&276.5KB	&276.5KB \cr
			\emph{E-FedDBL}		&55.4KB		&55.4KB \cr
			\emph{C-FedDBL} 	&55.4KB		&55.4KB	\cr
			\bottomrule
			\hspace{1mm}
		\end{tabular*}
	\end{threeparttable}
	\label{tab:Overhead}
\end{table}

\subsection{Comparisons under Multiple-round Communication}
\label{sub:multi-round}
In this experiment, we compare \emph{FedDBL} under one-round communication with centralized learning and two federated frameworks under multiple-round communication. Table~\ref{tab:mcc} demonstrates the average MCC performance. As expected, the centralized learning strategy achieves the best classification performance when with enough training data in MC-CRC dataset. \emph{C-FedDBL} with domain-relevant feature extractor constantly surpasses two federated learning frameworks, and even outperforms centralized learning when with less than 10\% training data. In BCSS, since this dataset is much smaller than MC-CRC, even \emph{FedDBL} with ImageNet pre-trained feature extractors \textit{R-} and \textit{E-} can outperform both centralized learning and existing federated frameworks and \emph{C-FedDBL} can further improve the quantitative results to a remarkable level.

\begin{figure*}[htp]
	\setlength{\abovecaptionskip}{0.cm}
	\setlength{\belowcaptionskip}{-0.cm}
	\centering
	\begin{tabular}{>{\centering\arraybackslash\hspace{0pt}}p{\linewidth}}
		\includegraphics[width=\linewidth]{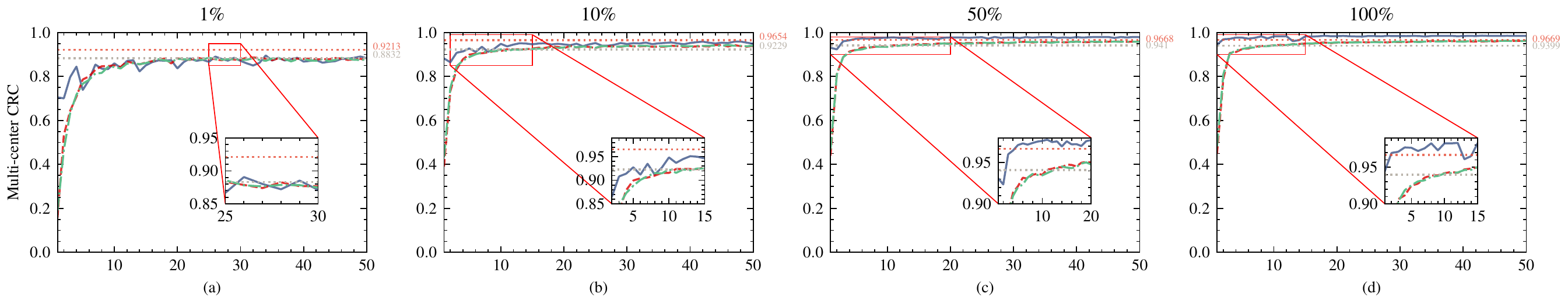}
		\includegraphics[width=\linewidth]{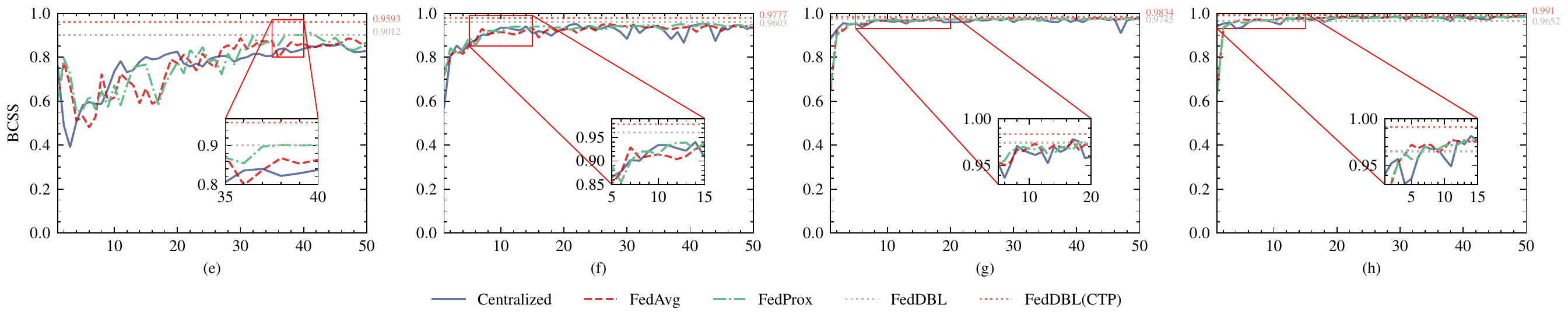}
	\end{tabular}
	\caption{Average global accuracy scores of 5-fold cross-validation results at each training epoch. Four representative dataset proportions are selected for visualization. Since \emph{FedDBL} is a one-round communication framework, we show \emph{FedDBL} and \emph{FedDBL(CTP)} results by a gray dash line and an orange dash line, respectively. We also highlight the most representative region for each sub-figure for better visualization.}
	\label{fig:Epoch_plot}
\end{figure*}

Moreover, we also visualize every epochs' average accuracy during training in Fig.~\ref{fig:Epoch_plot} to demonstrate the convergence speed of the existing approaches trained with four representative dataset proportions. Since \emph{FedDBL} is a one-round communication framework, we show \emph{R-FedDBL} and \emph{C-FedDBL} results by a gray dash line and an orange dash line, respectively. The most representative region in each sub-figure has been highlighted by a zoom-in window. As we can see, the convergence speed and the optimal performance of the existing models highly depend on the proportion of the training data. When with 100\% training data in MC-CRC, centralized learning can fast surpass \emph{R-FedDBL} within five epochs and even outperform \emph{C-FedDBL}. Two federated frameworks need more training epochs to converge and can achieve comparable performance with \emph{C-FedDBL}. When reducing the training samples, the convergence speed is becoming slower and the optimal performance also decreases. Particularly for 1\% training samples in BCSS, three existing models vibrate heavily and are even not able to surpass \emph{R-FedDBL} within 50 training epochs.

All the above experimental results have demonstrated the data efficiency, communication efficiency, model flexibility and model robustness of \emph{FedDBL} on histopathological image classification. \emph{FedDBL}, in our opinion, has the potential to significantly save computational and communication resources, relieve the pathologist's labeling burden and preserve the patient's privacy, which greatly promotes its clinical practicability compared with existing approaches.

\section{Scalability of FedDBL}

\subsection{Validation on External Dataset}
Model generalization is another criterion to evaluate the robustness on the unseen data. Following the same experimental settings of Table~\ref{tab:mcc}, we report MCC on the Kather Multiclass External (KME) dataset which is an external subset of Kather~\cite{kather2019predicting} and contains 7180 patches. Since \emph{FedDBL} greatly outperforms other federated training strategies under one-round communication. We train the centralized learning models and FL competitors under the setting of 50-round communication, to ensure they have learned a robust feature representation.
Table~\ref{tab:external-MCC} shows the average MCC on the KME dataset. When equipped with domain-irrelevant backbones, ResNet-50 and EfficientViT pre-trained on ImageNet, existing FL models (with 50-round communication) outperform \emph{FedDBL} (with one-round communication) with no doubt. However, their communication overhead is tens of thousands of times higher than \emph{FedDBL}. Even with only one-round communication, \emph{FedDBL} is still able to achieve moderate classification results. When we equip a domain-relevant DL-module (CTransPath), \emph{C-FedDBL} (with one-round communication) outperforms other competitors (with one-round communication) in most of the data proportions, which justifies the flexibility, communication and data efficiency of \emph{FedDBL}.
\begin{table*}[th]
	\centering
	\caption{External validation on KME trained under MC-CRC (MCC). Each value is calculated with five-fold cross-validation experiments. For simplification, the detail settings can be found in Table~\ref{tab:CRC_one-round-mcc}. $\dagger$ means models trained for 50 rounds.}
	\begin{threeparttable}
		\setlength{\tabcolsep}{2.5mm}{
			\begin{tabular}{
					p{0.09\textwidth}>{\centering}
					p{0.09\textwidth}>{\centering}
					p{0.09\textwidth}>{\centering}
					p{0.09\textwidth}>{\centering}
					p{0.09\textwidth}>{\centering}
					p{0.09\textwidth}>{\centering}
					p{0.09\textwidth}>{\centering}
					p{0.09\textwidth}}
				\toprule
				\multirow{2}{*}{Models} & \multicolumn{7}{c}{MCC of External KME under Multiple-round Training} \cr 
				\cmidrule(lr){2-8} 
				&1\% &5\% &10\% &30\% &50\% &70\% &100\%  \cr
				\midrule
					\multirow{2}{*}[1.2ex]{\emph{R-Centralized$\dagger$}} 	&0.8738 \\ (0.872, 0.874)
																		&0.9191 \\ (0.918, 0.920)
																		&0.9352 \\ (0.935, 0.936)
																		&0.9403 \\ (0.940, 0.941)
																		&0.9487 \\ (0.948, 0.949)
																		&0.9555 \\ (0.955, 0.956)
																		&0.9509 \\ (0.950, 0.951)	\cr
					\multirow{2}{*}[1.2ex]{\emph{E-Centralized$\dagger$}} 	&0.8842 \\ (0.884, 0.887)
																		&0.9208 \\ (0.920, 0.921)
																		&0.9437 \\ (0.943, 0.944)
																		&0.9527 \\ (0.952, 0.953)
																		&\textcolor{red}{0.9512} \\ (0.951, 0.951)
																		&\textcolor{red}{0.9586} \\ (0.958, 0.959)
																		&\textcolor{red}{0.9568} \\ (0.957, 0.957)	\cr
					\cdashline{1-8}
					\multirow{2}{*}[1.2ex]{\emph{R-FedAvg$\dagger$}}	&\textbf{\textcolor{red}{0.8983}} \\ (0.898, 0.899)
																&0.9330 \\ (0.933, 0.934)
																&0.9350 \\ (0.935, 0.936)
																&0.9392 \\ (0.939, 0.940)
																&0.9387 \\ (0.939, 0.940)
																&0.9382 \\ (0.938, 0.938)
																&0.9329 \\ (0.932, 0.933)	\cr
					\multirow{2}{*}[1.2ex]{\emph{R-FedProx$\dagger$}}     &0.8833 \\ (0.882, 0.884)
																	&0.9309 \\ (0.930, 0.932)
																	&0.9320 \\ (0.931, 0.932)
																	&0.9398 \\ (0.939, 0.940)
																	&0.9384 \\ (0.938, 0.939)
																	&0.9332 \\ (0.933, 0.933)
																	&0.9344 \\ (0.934, 0.935)	\cr
					\multirow{2}{*}[1.2ex]{\emph{E-FedAvg$\dagger$}}  &0.8958 \\ (0.895, 0.898)
																&0.9328 \\ (0.932, 0.933)
																&0.9436 \\ (0.943, 0.945)
																&0.9427 \\ (0.942, 0.943)
																&0.9449 \\ (0.944, 0.945)
																&0.9395 \\ (0.939, 0.940)
																&0.9350 \\ (0.934, 0.935)	\cr
					\multirow{2}{*}[1.2ex]{\emph{E-FedProx$\dagger$}}		&0.8974 \\ (0.895, 0.898)
																	&0.9336 \\ (0.933, 0.935)
																	&0.9339 \\ (0.934, 0.935)
																	&\textbf{\textcolor{red}{0.9501}} \\ (0.950, 0.951)
																	&0.9438 \\ (0.943, 0.944)
																	&0.9430 \\ (0.943, 0.944)
																	&0.9449 \\ (0.944, 0.945)	\cr

					\multirow{2}{*}[1.2ex]{\emph{R-FedDBL(ours)}}	&0.8280 \\ (0.827, 0.829)
														&0.6796 \\ (0.676, 0.681)
														&0.8315 \\ (0.830, 0.833)
														&0.8852 \\ (0.884, 0.886)
														&0.8928 \\ (0.893, 0.894)
														&0.8942 \\ (0.894, 0.895)
														&0.8951 \\ (0.895, 0.896)	\cr
					\multirow{2}{*}[1.2ex]{\emph{E-FedDBL(ours)}}	&0.8066 \\ (0.806, 0.807)
														&0.8424 \\ (0.840, 0.844)
														&0.8652 \\ (0.864, 0.867)
														&0.8938 \\ (0.893, 0.894)
														&0.8855 \\ (0.885, 0.886)
														&0.8905 \\ (0.890, 0.891)
														&0.8931 \\ (0.893, 0.893)	\cr
					\multirow{2}{*}[1.2ex]{\emph{C-FedDBL(ours)}}  	&0.8615 \\ (0.861, 0.864)
														&\textbf{\textcolor{red}{0.9360}} \\ (0.936, 0.937)
														&\textbf{\textcolor{red}{0.9445}} \\ (0.944, 0.945)
														&0.9463 \\ (0.946, 0.946)
														&\textbf{0.9475} \\ (0.947, 0.948)
														&\textbf{0.9476} \\ (0.947, 0.948)
														&\textbf{0.9471} \\ (0.947, 0.948)	\cr
				\bottomrule
				\hspace{1mm}
		\end{tabular}}
	\end{threeparttable}
	\label{tab:external-MCC}
\end{table*}

\subsection{Different Number of Clients}
In this experiment, we aim to evaluate the model performance of \emph{FedDBL} with different numbers of clients. In order to avoid the independent and identically distributed (IID) data problem, we let the original clients on the MC-CRC dataset as ParentClients and divide each of them into ChildClients, where $n\in\{5,10,15,20\}$. The total number of clients is now $\{20,40,60,80\}$. This experiment runs on 10\%, 50\%, and 100\% training samples. The reason we do not reduce the training samples to 1\% is to maintain sufficient training samples for each client since increasing the number of clients will decrease the number of training samples.

\begin{figure}[t]
	\centering
	\includegraphics[width=0.975\linewidth]{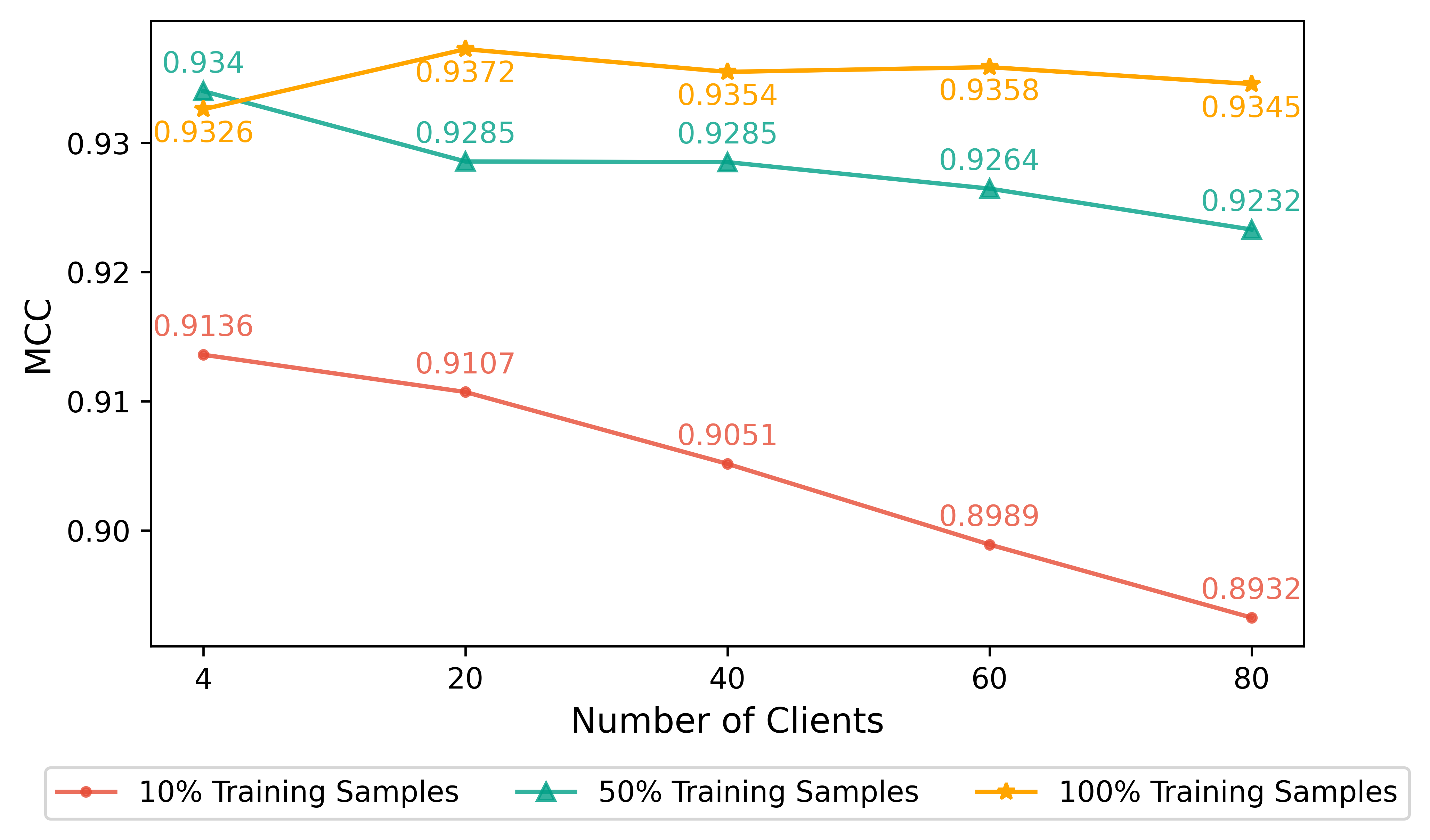}
	\caption{Average MCC of 5-fold cross-validation results on MC-CRC with different number of clients. \emph{FedDBL} with ResNet-50 is the used here. Each line indicates \emph{R-FedDBL} trained on different proportions of training samples.}
	\label{fig:multi_clients_plot}
\end{figure}

As demonstrated in Fig.~\ref{fig:multi_clients_plot}, when with small-scale training samples, i.e., 10\% and 50\%, the classification performance slightly drops with the increasing number of clients as we expected. Because the total number of training samples is fixed, more clients mean fewer training samples for each client. When the experiment runs on 100\% training samples, the results become more stable even when we increase the number of clients to 80. According to the experimental results, \emph{FedDBL} is robust to different numbers of clients.

\begin{table}[t]
	\centering
	\caption{Experiment on model personalization. F1-score is reported on MC-CRC on each client with 1\% training samples. The baseline in this experiment is \emph{C-FedDBL}. Local means each specific client trains its model on the private data locally. Global and personalized mean different model aggregation weights. We validate the models on both local test sets and the external test set (KME).}
	\begin{threeparttable}
		\begin{tabular}{lcccc}
			\toprule
			Models & \#1 & \#2 & \#3 & \#4  \cr
			\cmidrule(lr){1-5}			
			 \multicolumn{5}{c}{\textbf{Validation on Local Test Set}} \cr
			\emph{C-FedDBL} (Local)				&0.8960 	&0.9072 	&0.9437 	&0.9778   \cr
			\emph{C-FedDBL} (Global) 			&0.9071 	&0.8942 	&0.9526 	&0.9534   \cr
			\emph{C-FedDBL} (Personalized)     &0.9383 	&0.9369 &0.9652 &0.9812   \cr

			\midrule
			 \multicolumn{5}{c}{\textbf{Validation on External Test Set (KME)}} \cr 
			\emph{C-FedDBL} (Local)				&0.5431 	&0.8031 	&0.5881 	&0.7404   \cr
			\emph{C-FedDBL} (Global) 			&\multicolumn{4}{c}{0.8379}  \cr
			\emph{C-FedDBL} (Personalized)     &0.7508 	&0.8725 	&0.7827 	&0.8348    \cr

			\bottomrule
			\hspace{1mm}
		\end{tabular}
	\end{threeparttable}
	\label{tab:personalization_acc_f1}
\end{table} 
\subsection{Model Personalization}
Personalized Federated Learning (PFL) is an important topic in the FL domain. However, conventional DL-based federated frameworks rely on specific designs to achieve PFL, such as knowledge distillation~\cite{Chen2023ICLR}, parameter decoupling~\cite{Dong2023SplitGP} and etc. Thanks to the simple architecture, \emph{FedDBL} can easily achieve model personalization by adjusting the model contributions of the global model and the local model by Eq.~\ref{eq:personalization}.
\begin{equation}
	W_{personalized}^k = \lambda \times W_{local}^k + (1-\lambda) \times W_{global}
	\label{eq:personalization}
\end{equation}
where $W_{local}^k$ is the local model of the $k$-th client and $W_{global}$ is the global model. $\lambda$ controls the contribution of the global and local models. In this experiment, we compare three different \emph{C-FedDBL} baselines. (1) \emph{C-FedDBL} (Local): When $\lambda=1$, each client trains its own data locally. (2) \emph{C-FedDBL} (Global): When $\lambda=0$, all the clients use the same global model. (3) \emph{C-FedDBL} (Personalized): When $\lambda=0.5$, each client achieves personalization by weighted aggregating the global model. The experiment runs on 1\% training samples. As demonstrated in Table~\ref{tab:personalization_acc_f1}, when validating the models on the local test sets, the personalized models show better performance compared with local and global models. Because aggregating other local models can increase the total number of training samples when with a small-scale training set. However, when validating the models on the external dataset KME, the local model is strongly affected by the domain gap between the local data and the external data. Personalized models can alleviate the domain gap but are still not generalizable. The global model achieves the best performance on KME with the best model generalization.

\begin{table}[t]
	\centering
	\caption{Comparison between \emph{C-FedDBL} and \emph{C-FedDBL(encrypted)} under 1\% MC-CRC and CTransPath backbone.}
	\begin{threeparttable}
		\begin{tabular}{ccccc}
			\toprule
			Models 						&\makecell[c]{Accuracy}	&\makecell[c]{F1-score}	&\makecell[c]{MCC} &\makecell[c]{Model size} 	\cr
			\midrule
			\makecell[c]{\emph{C-FedDBL}} &\makecell[c]{0.9213}	&\makecell[c]{0.9217}	&\makecell[c]{0.9114}&\makecell[c]{55.4KB}		\cr
			\makecell[c]{\emph{C-FedDBL} \\ \emph{(encrypted)}}	&\makecell[c]{0.9213}	&\makecell[c]{0.9217}	&\makecell[c]{0.9114} &\makecell[c]{30MB}			\cr
			\bottomrule
			\hspace{1mm}
		\end{tabular}
	\end{threeparttable}
	\label{tab:encrypted}
\end{table}
\subsection{Privacy-preserving with Encryption}
One of the objectives of \emph{FedDBL} is to preserve patients' privacy. However, even though it is not necessary to directly share the raw data, conventional DL-based federated frameworks still suffer from different kinds of federated attacks, such as model inversion attack~\cite{zhu2019deep} or man-in-the-middle attack~\cite{wang2020man}. Compared with conventional DL-based federated frameworks, \emph{FedDBL} can defend against the aforementioned attacks because the training data is totally unseen for the DL-module and it is the deep features that are used for calculating the parameters of the BL-module at each client locally. Actually, the pre-trained DL-module can be regarded as a perturbation process to transfer the data into high-level features. And the features are also protected by only sharing the parameters of the BL-module.

Thanks to the lightweight BL-module, we can further protect the parameters by employing an additional encryption algorithm~\cite{aono2017efficient} to support federated aggregation in the encrypted domain. The corresponding packing-encryption~\cite{le2018privacy} is used where it exploits the redundancy of ciphertext space to hold more plaintext in an encrypted block. Table~\ref{tab:encrypted} demonstrates the model accuracy, F1-score, MCC and the model size of the packing-encrypted \emph{C-FedDBL} with 1\% MC-CRC dataset using CTransPath pre-trained backbone. When the bit-length for the encryption precision is set as 32 bits, the encryption algorithm does not harm the parameters. So the model accuracy is preserved. By using packing-encryption, the model size is still much smaller than that of ResNet-50 in the plaintext condition (94.4MB) shown in Table~\ref{tab:Overhead}.

\begin{table}[t]
	\centering
	\caption{Experiment on other image modality. MCC on OrganACS with different proportions of the training data using ResNet-50 architecture. \red{Red} represents the best results under the one-round training configuration.}
	\begin{threeparttable}
		\setlength{\tabcolsep}{2.5mm}{
			\begin{tabular}{lcccc}
				\toprule
				\multirow{2}{*}{Models} &\multirow{2}{*}{Rounds} &\multicolumn{3}{c}{MCC on OrganACS} \cr 
				\cmidrule(lr){3-5} 
				& &1\% &50\% &100\%  \cr
				\midrule
				\emph{R-Centralized} 	&1	&0.3201	&0.7724	&\textcolor{red}{0.8178} \cr
				\emph{R-FedAvg} 		&1 	&0.1282	&0.1744	&0.1659 \cr
				\emph{R-FedProx}     	&1	&0.1326	&0.1191	&0.1177 \cr

				\emph{R-FedDBL(ours)} 	&1	&\textcolor{red}{0.6095}	&\textcolor{red}{0.8042}	&0.8123\cr

				\cdashline{1-5}
				\emph{R-Centralized} 	&50	&0.5841	&0.8491	&0.8642\cr
				\emph{R-FedAvg} 		&50	&0.6397	&0.8131	&0.8290\cr
				\emph{R-FedProx}     	&50	&0.6216	&0.8117	&0.8094\cr

				\bottomrule
				\hspace{1mm}
		\end{tabular}}
	\end{threeparttable}
	\label{tab:organ-mcc}
\end{table}
\subsection{Other Image Modality}
To test the extendibility of \emph{FedDBL} on the other medical image modality, we conduct an additional experiment on an abdominal CT dataset, named organ\{A, C, S\}MNIST~\cite{medmnistv1,medmnistv2} (in short OrganACS), for 11 body organs classification. This dataset consists of 107,731 cropped 2D images in Axial / Coronal / Sagittal views. This experiment runs on 1\%, 50\% and 100\% training samples. As demonstrated in Table~\ref{tab:organ-mcc}, with 100\% training samples and sufficient communication budgets, \emph{R-FedDBL} and the other two conventional federated frameworks achieve comparable results, but are still less effective than centralized learning. When with insufficient communication budgets, \emph{R-FedAvg} and \emph{R-FedProx} become unstable and fail to provide favorable results. Because the local models are hard to converge with only one-round local training. Merging the unstable local models might further harm the feature representation of the global model. Similar to the conclusion on MC-CRC and BCSS, \emph{R-FedDBL} is the best solution for the conditions of limited training samples and communication budgets.
\section{Discussion and Conclusion}
In this paper, we proposed a novel federated framework (FedDBL) for histopathological tissue classification. Thanks to the robust deep learning feature extractor and flexible broad learning inference system, FedDBL can greatly improve the classification performance with only one-round communication and extremely limited training samples, which significantly reduces the data dependency and communication burden. With the employment of DL-module and BL-module for local training, each client just needs to solve a lightweight broad learning system, which drastically reduces the training overhead. Sharing the lightweight BL-module in the federated framework not only greatly saves the communication burden, but also preserves the patients' privacy. Experimental results with five-fold cross-validation demonstrate that FedDBL outperforms both centralized learning strategies and federated frameworks in one-round communication. It even outperforms all the competitors in 50-round training iterations when with limited training samples. Addition experiments show the scalability of FedDBL on different number of clients, model personalization, privacy-preserving with encryption and the other medical image modality.

Moreover, due to the flexible model design, FedDBL can be further upgraded by replacing any module with a superior one in the future. In this paper, we have proven that a domain-relevant deep feature extractor is more effective than a domain-irrelevant one. Since the federated framework in this study is the most common federated average aggregation strategy. We expect a more outstanding federated aggregation framework can be applied in the future to further improve the performance of FedDBL under extreme data and communication situations.

\section*{Acknowledgments}
This work was supported by Key-Area Research and Development Program of Guangdong Province (No. 2021B0101420006), 
Regional Innovation and Development Joint Fund of National Natural Science Foundation of China (No. U22A20345), 
National Science Foundation for Young Scientists of China (No. 62102103),
Natural Science Foundation for Distinguished Young Scholars of Guangdong Province (No. 2023B1515020043).

\bibliography{refer}
\bibliographystyle{IEEEtran}
\newpage

\begin{table*}[t]
	\centering
	\caption{Average accuracy on MC-CRC dataset with different proportions of the training data under one-round training. For simplification, the detail settings can be found in Table~\ref{tab:CRC_one-round-mcc}.}
	\begin{threeparttable}
		\setlength{\tabcolsep}{2.5mm}{
			\begin{tabular}{
					p{0.09\textwidth}>{\centering}
					p{0.09\textwidth}>{\centering}
					p{0.09\textwidth}>{\centering}
					p{0.09\textwidth}>{\centering}
					p{0.09\textwidth}>{\centering}
					p{0.09\textwidth}>{\centering}
					p{0.09\textwidth}>{\centering}
					p{0.09\textwidth}}
				\toprule
				\multirow{2}{*}{Models} & \multicolumn{7}{c}{Accuracy under One-round Training on MC-CRC} \cr 
				\cmidrule(lr){2-8} 
				&	1\% &	5\% &	10\% &	30\% &	50\% &	70\% &	100\%  \cr
				\midrule
				\multirow{2}{*}[1.2ex]{\emph{R-Centralized}} 	&0.7185 \\ (0.713, 0.720)
																&0.7913 \\ (0.791, 0.799)
																&0.8819 \\ (0.881, 0.884)
																&0.9033 \\ (0.903, 0.905)
																&0.9306 \\ (0.93, 0.931)
																&\textcolor{red}{0.9414}\\ (0.941, 0.942)
																&0.9450 \\ (0.943, 0.945)	\cr
				\multirow{2}{*}[1.2ex]{\emph{R-Centralized-FC}}   &0.8334 \\ (0.833, 0.835)
																	&\textcolor{red}{0.8970} \\ (0.897, 0.897)
																	&0.9129 \\ (0.912, 0.913)
																	&0.9260 \\ (0.925, 0.926)
																	&0.9378 \\ (0.937, 0.938)
																	&0.9412 \\ (0.941, 0.942)
																	&\textcolor{red}{0.9458} \\ (0.945, 0.946)	\cr
				\multirow{2}{*}[1.2ex]{\emph{R-FedAvg}}   &0.1465	\\ (0.145, 0.148)
														&0.3095	\\ (0.303, 0.315)
														&0.4025	\\ (0.400, 0.408)
														&0.4145	\\ (0.409, 0.417)
														&0.4461	\\ (0.444, 0.448)
														&0.4823	\\ (0.482, 0.486)
														&0.4345 \\ (0.431, 0.435)	\cr
				\multirow{2}{*}[1.2ex]{\emph{R-FedProx}}    &0.1900	\\ (0.189, 0.197)
															&0.2976	\\ (0.294, 0.301)
															&0.4332	\\ (0.432, 0.437)
															&0.4621	\\ (0.462, 0.466)
															&0.4663	\\ (0.463, 0.468)
															&0.4448	\\ (0.444, 0.446)
															&0.4818 \\ (0.480, 0.486)	\cr
				\multirow{2}{*}[1.2ex]{\emph{R-FedPAQ}}	&0.1984	\\ (0.195, 0.200)
														&0.3373	\\ (0.336, 0.341)
														&0.3798	\\ (0.377, 0.382)
														&0.5208	\\ (0.519, 0.523)
														&0.4778	\\ (0.475, 0.479)
														&0.4919	\\ (0.491, 0.495)
														&0.4994 \\ (0.498, 0.504)	\cr
				\multirow{2}{*}[1.2ex]{\emph{R-FedAvg-FC}}   &0.7942 \\ (0.794, 0.795)
																&\textbf{0.8552} \\ (0.855, 0.856)
																&0.8628	\\ (0.862, 0.863)
																&0.8817	\\ (0.881, 0.882)
																&0.8900	\\ (0.890, 0.891)
																&0.8857	\\ (0.885, 0.886)
																&0.8977 \\ (0.897, 0.898)	\cr
				\multirow{2}{*}[1.2ex]{\emph{R-FedDBL(ours)}}  &\textbf{\textcolor{red}{0.8832}} \\ (0.883, 0.884)
																	&0.8456	\\ (0.845, 0.846)
																	&\textbf{\textcolor{red}{0.9229}} \\ (0.922, 0.923)
																	&\textbf{\textcolor{red}{0.9411}} \\ (0.941, 0.941)
																	&\textbf{\textcolor{red}{0.9410}} \\ (0.941, 0.941)
																	&\textbf{0.9413} \\ (0.941, 0.942)
																	&\textbf{0.9399} \\ (0.939, 0.940)	 \cr
				
				\cdashline{1-8}
				\multirow{2}{*}[1.2ex]{\emph{E-Centralized}} 	&0.8001 \\ (0.799, 0.802)
																&0.8996 \\ (0.898, 0.901)
																&0.9190 \\ (0.917, 0.919)
																&\textcolor{red}{0.9468} \\ (0.947, 0.947)
																&\textcolor{red}{0.9490} \\ (0.947, 0.950)
																&\textcolor{red}{0.9587} \\ (0.959, 0.959)
																&\textcolor{red}{0.9614} \\ (0.961, 0.962)	\cr
				\multirow{2}{*}[1.2ex]{\emph{E-Centralized-FC}}     &0.8077 \\ (0.807, 0.808)
																	&0.8795 \\ (0.879, 0.880)
																	&0.9003 \\ (0.899, 0.901)
																	&0.9243 \\ (0.924, 0.925)
																	&0.9318 \\ (0.931, 0.932)
																	&0.9345 \\ (0.934, 0.935)
																	&0.9370 \\ (0.936, 0.937)	\cr
				\multirow{2}{*}[1.2ex]{\emph{E-FedAvg}}      	&0.6820 \\ (0.678, 0.687)
																&0.7876 \\ (0.786, 0.789)
																&0.7906 \\ (0.790, 0.793)
																&0.8178 \\ (0.817, 0.821)
																&0.8099 \\ (0.808, 0.811)
																&0.8183 \\ (0.818, 0.819)
																&0.8227 \\ (0.822, 0.823)	\cr
				\multirow{2}{*}[1.2ex]{\emph{E-FedProx}}     	&0.7166 \\ (0.714, 0.719)
																&0.7879 \\ (0.783, 0.789)
																&0.7683 \\ (0.766, 0.769)
																&0.7939 \\ (0.792, 0.796)
																&0.8181 \\ (0.817, 0.818)
																&0.8108 \\ (0.810, 0.813)
																&0.8211 \\ (0.820, 0.823)	\cr
				\multirow{2}{*}[1.2ex]{\emph{E-FedPAQ}}		&0.6759 \\ (0.672, 0.678)
															&0.7878 \\ (0.786, 0.790)
															&0.7801 \\ (0.778, 0.785)
															&0.8081 \\ (0.806, 0.808)
															&0.8348 \\ (0.833, 0.836)
															&0.8259 \\ (0.826, 0.829)
															&0.8175 \\ (0.816, 0.819)	\cr
				\multirow{2}{*}[1.2ex]{\emph{E-FedAvg-FC}}   	&0.7190 \\ (0.718, 0.720)
																&0.8298 \\ (0.829, 0.830)
																&0.8535 \\ (0.853, 0.854)
																&0.8821 \\ (0.882, 0.883)
																&0.8916 \\ (0.891, 0.892)
																&0.8977 \\ (0.897, 0.898)
																&0.9013 \\ (0.901, 0.901)	\cr
				\multirow{2}{*}[1.2ex]{\emph{E-FedDBL(ours)}}  	&\textcolor{red}{\textbf{0.8515}} \\ (0.851, 0.852)
																	&\textcolor{red}{\textbf{0.9132}} \\ (0.913, 0.914)
																	&\textcolor{red}{\textbf{0.9275}} \\ (0.927, 0.928)
																	&\textbf{0.9331} \\ (0.932, 0.933)
																	&\textbf{0.9329} \\ (0.932, 0.933)
																	&\textbf{0.9328} \\ (0.932, 0.933)
																	&\textbf{0.9330} \\ (0.932, 0.933)	\cr
				
				\cdashline{1-8}
				\multirow{2}{*}[1.2ex]{\emph{C-Centralized-FC}}	&\textcolor{red}{0.9390} \\ (0.939, 0.939)
															&0.9594 \\ (0.959, 0.959)
															&\textcolor{red}{0.9670} \\ (0.967, 0.967)
															&\textcolor{red}{0.9756} \\ (0.975, 0.976)
															&\textcolor{red}{0.9788} \\ (0.978, 0.979)
															&\textcolor{red}{0.9801} \\ (0.980, 0.981)
															&\textcolor{red}{0.9817} \\ (0.981, 0.982)	\cr
				\multirow{2}{*}[1.2ex]{\emph{C-FedAvg-FC}}	  	&0.9074	\\ (0.907, 0.908)
															&0.9382 \\ (0.938, 0.939)
															&0.9455 \\ (0.945, 0.946)
															&0.9536 \\ (0.953, 0.954)
															&0.9563 \\ (0.956, 0.956)
															&0.9577 \\ (0.957, 0.958)
															&0.9595 \\ (0.959, 0.960)	\cr
				\multirow{2}{*}[1.2ex]{\emph{C-FedDBL(ours)}}  &\textbf{0.9213} \\ (0.921, 0.922)
															&\textbf{\textcolor{red}{0.9640}} \\ (0.964, 0.964)
															&\textbf{0.9654} \\ (0.965, 0.966)
															&\textbf{0.9663} \\ (0.966, 0.967)
															&\textbf{0.9668} \\ (0.966, 0.967)
															&\textbf{0.9669} \\ (0.966, 0.967)
															&\textbf{0.9669} \\ (0.966, 0.967)	\cr
				
				\bottomrule
				\hspace{1mm}
		\end{tabular}}
	\end{threeparttable}
	\label{tab:CRC_one-round-accuracy}
\end{table*}

\begin{table*}[t]
	\centering
	\caption{Average accuracy on BCSS dataset with different proportions of the training data under one-round training. For simplification, the detail settings can be found in Table~\ref{tab:CRC_one-round-mcc}.}
	\begin{threeparttable}
		\setlength{\tabcolsep}{2.5mm}{
			\begin{tabular}{p{0.09\textwidth}>{\centering}
					p{0.09\textwidth}>{\centering}
					p{0.09\textwidth}>{\centering}
					p{0.09\textwidth}>{\centering}
					p{0.09\textwidth}>{\centering}
					p{0.09\textwidth}>{\centering}
					p{0.09\textwidth}>{\centering}
					p{0.09\textwidth}}
				\toprule
				\multirow{2}{*}{Models} & \multicolumn{7}{c}{Accuracy under One-round Training on BCSS} \cr 
				\cmidrule(lr){2-8} 
				&1\% &5\% &10\% &30\% &50\% &70\% &100\%  \cr
				\midrule
				\multirow{2}{*}[1.2ex]{\emph{R-Centralized}}	&0.7654 \\ (0.760, 0.769)
														&0.4843 \\ (0.471, 0.488)
														&0.5879 \\ (0.581, 0.601)
														&0.8757 \\ (0.875, 0.887)
														&0.8888 \\ (0.888, 0.891)
														&0.9033 \\ (0.900, 0.905)	
														&0.9392 \\ (0.938, 0.939)	\cr
				\multirow{2}{*}[1.2ex]{\emph{R-Centralized-FC}} &0.5806 \\ (0.580, 0.582)
														&0.8106	\\ (0.810, 0.812)
														&0.8640	\\ (0.863, 0.865)
														&0.9345	\\ (0.934, 0.935)
														&0.9490	\\ (0.948, 0.949)
														&0.9543	\\ (0.954, 0.955)
														&0.9600 \\ (0.959, 0.960)	\cr
				\multirow{2}{*}[1.2ex]{\emph{R-FedAvg}}	&0.5972 \\ (0.594, 0.598)
													&0.7495 \\ (0.749, 0.752)
													&0.7062 \\ (0.706, 0.712)
													&0.6959 \\ (0.689, 0.698)
													&0.6611 \\ (0.658, 0.669)
													&0.5889 \\ (0.587, 0.591)
													&0.6420 \\ (0.638, 0.649)	\cr
				\multirow{2}{*}[1.2ex]{\emph{R-FedProx}}	&0.5951 \\ (0.592, 0.596)
													&0.7277 \\ (0.725, 0.730)
													&0.7158 \\ (0.709, 0.720)	
													&0.7155 \\ (0.709, 0.718)
													&0.6654 \\ (0.663, 0.673)
													&0.6631 \\ (0.655, 0.664)
													&0.6308 \\ (0.622, 0.632)	\cr
				\multirow{2}{*}[1.2ex]{\emph{R-FedPAQ}}	&0.5929	\\ (0.589, 0.593)
													&0.6679	\\ (0.660, 0.670)
													&0.6592	\\ (0.653, 0.663)
													&0.6776	\\ (0.677, 0.681)
													&0.6025	\\ (0.601, 0.604)
													&0.6812	\\ (0.673, 0.684)
													&0.5374 \\ (0.530, 0.546)	\cr
				\multirow{2}{*}[1.2ex]{\emph{R-FedAvg-FC}}	&0.5740 \\ (0.573, 0.575)
													&0.6234 \\ (0.619, 0.624)	
													&0.7714 \\ (0.770, 0.773)
													&0.8754 \\ (0.875, 0.876)
													&0.9014 \\ (0.901, 0.902)
													&0.9259 \\ (0.926, 0.926)
													&0.9365 \\ (0.936, 0.937)	\cr
				\multirow{2}{*}[1.2ex]{\emph{R-FedDBL(ours)}}	&\textbf{\textcolor{red}{0.9012}} \\ (0.900, 0.902)
															&\textbf{\textcolor{red}{0.9511}} \\ (0.951, 0.951)
															&\textbf{\textcolor{red}{0.9603}} \\ (0.960, 0.961)
															&\textbf{\textcolor{red}{0.9711}} \\ (0.971, 0.971)
															&\textbf{\textcolor{red}{0.9745}} \\ (0.974, 0.975)
															&\textbf{\textcolor{red}{0.9731}} \\ (0.973, 0.973)
															&\textbf{\textcolor{red}{0.9652}} \\ (0.965, 0.965)	\cr
				
				\cdashline{1-8}
				\multirow{2}{*}[1.2ex]{\emph{E-Centralized}} 	&0.5384 \\ (0.536, 0.544)
																&0.8444 \\ (0.839, 0.847)
																&0.9098 \\ (0.908, 0.911)
																&0.8791 \\ (0.876, 0.882)
																&0.9298 \\ (0.927, 0.930)
																&0.9350 \\ (0.932, 0.936)
																&0.9448 \\ (0.943, 0.946)	\cr
				\multirow{2}{*}[1.2ex]{\emph{E-Centralized-FC}}     &0.6241 \\ (0.621, 0.624)
																	&0.6188 \\ (0.617, 0.619)
																	&0.7845 \\ (0.784, 0.786)
																	&0.8777 \\ (0.877, 0.878)
																	&0.9127 \\ (0.912, 0.913)
																	&0.9344 \\ (0.934, 0.935)
																	&0.9458 \\ (0.945, 0.946)	\cr
				\multirow{2}{*}[1.2ex]{\emph{E-FedAvg}}     &0.2230 \\ (0.217, 0.230)
															&0.5790 \\ (0.571, 0.587)
															&0.7258 \\ (0.721, 0.737)
															&0.9364 \\ (0.935, 0.937)
															&0.9327 \\ (0.931, 0.933)
															&0.9449 \\ (0.945, 0.946)
															&0.9384 \\ (0.938, 0.940)	\cr
				\multirow{2}{*}[1.2ex]{\emph{E-FedProx}}     	&0.2266 \\ (0.221, 0.233)
																&0.6121 \\ (0.603, 0.616)
																&0.7879 \\ (0.787, 0.795)
																&0.9354 \\ (0.934, 0.936)
																&0.9324 \\ (0.931, 0.934)
																&0.9447 \\ (0.943, 0.945)
																&0.9350 \\ (0.935, 0.936)	\cr
				\multirow{2}{*}[1.2ex]{\emph{E-FedPAQ}}		&0.3283 \\ (0.318, 0.336)
															&0.5841 \\ (0.581, 0.591)
															&0.7843 \\ (0.775, 0.787)
															&0.9406 \\ (0.940, 0.941)
															&0.9397 \\ (0.939, 0.940)
															&0.9444 \\ (0.943, 0.945)
															&0.9368 \\ (0.936, 0.938)	\cr
				\multirow{2}{*}[1.2ex]{\emph{E-FedAvg-FC}}   	&0.4879 \\ (0.477, 0.494)
																&0.5966 \\ (0.595, 0.599)
																&0.5738 \\ (0.573, 0.574)
																&0.7883 \\ (0.787, 0.789)
																&0.8232 \\ (0.823, 0.823)
																&0.8601 \\ (0.860, 0.861)
																&0.8867 \\ (0.886, 0.887)	\cr
				\multirow{2}{*}[1.2ex]{\emph{E-FedDBL(ours)}}    &\textbf{\textcolor{red}{0.8971}} \\ (0.897, 0.898)
																	&\textbf{\textcolor{red}{0.9368}} \\ (0.936, 0.937)
																	&\textbf{\textcolor{red}{0.9324}} \\ (0.932, 0.933)
																	&\textbf{\textcolor{red}{0.9485}} \\ (0.948, 0.949)
																	&\textbf{\textcolor{red}{0.9607}} \\ (0.960, 0.961)
																	&\textbf{\textcolor{red}{0.9640}} \\ (0.964, 0.964)
																	&\textbf{\textcolor{red}{0.9656}} \\ (0.965, 0.966)	\cr

				\cdashline{1-8}
				\multirow{2}{*}[1.2ex]{\emph{C-Centralized-FC}}	&0.6159	\\ (0.614, 0.618)
															&0.5904 \\ (0.589, 0.592)
															&0.6692 \\ (0.668, 0.670)
															&0.8886 \\ (0.886, 0.887)
															&0.9435 \\ (0.943, 0.944)
															&0.9609 \\ (0.960, 0.961)
															&0.9726 \\ (0.972, 0.973)	\cr
				\multirow{2}{*}[1.2ex]{\emph{C-FedAvg-FC}} 		&0.5858 \\ (0.585, 0.587)
															&0.5772 \\ (0.577, 0.578)
															&0.5676 \\ (0.567, 0.568)	
															&0.6689 \\ (0.668, 0.669)
															&0.8072 \\ (0.807, 0.807)
															&0.8537 \\ (0.853, 0.854)
															&0.9072 \\ (0.907, 0.908)	\cr
				\multirow{2}{*}[1.2ex]{\emph{C-FedDBL(ours)}}	&\textbf{\textcolor{red}{0.9593}} \\ (0.959, 0.960)
															&\textbf{\textcolor{red}{0.9754}} \\ (0.975, 0.976)
															&\textbf{\textcolor{red}{0.9777}} \\ (0.977, 0.978)
															&\textbf{\textcolor{red}{0.9770}} \\ (0.977, 0.978)
															&\textbf{\textcolor{red}{0.9834}} \\ (0.983, 0.983)
															&\textbf{\textcolor{red}{0.9883}} \\ (0.988, 0.988)
															&\textbf{\textcolor{red}{0.9910}} \\ (0.991, 0.991)	\cr
				
				\bottomrule
				\hspace{1mm}
		\end{tabular}}
	\end{threeparttable}
	\label{tab:BCSS_one-round-accuracy}
\end{table*}

\begin{table*}[th]
	\centering
	\caption{Average F1-score on MC-CRC dataset with different proportions of the training data under one-round training. For simplification, the detail settings can be found in Table~\ref{tab:CRC_one-round-mcc}.}
	\begin{threeparttable}
		\setlength{\tabcolsep}{2.5mm}{
			\begin{tabular}{p{0.09\textwidth}>{\centering}
					p{0.09\textwidth}>{\centering}
					p{0.09\textwidth}>{\centering}
					p{0.09\textwidth}>{\centering}
					p{0.09\textwidth}>{\centering}
					p{0.09\textwidth}>{\centering}
					p{0.09\textwidth}>{\centering}
					p{0.09\textwidth}}
				\toprule
				\multirow{2}{*}{Models} & \multicolumn{7}{c}{F1-score under One-round Training on MC-CRC} \cr 
				\cmidrule(lr){2-8} 
				&1\% &5\% &10\% &30\% &50\% &70\% &100\%  \cr
				\midrule
				\multirow{2}{*}[1.2ex]{\emph{R-Centralized}} 	&0.7191	\\ (0.714, 0.721)
														&0.7863	\\ (0.786, 0.794)
														&0.8801	\\ (0.879, 0.883)
														&0.9027	\\ (0.902, 0.905)
														&0.9300	\\ (0.929, 0.930)
														&0.9415	\\ (0.941, 0.942)
														&0.9451	\\ (0.943, 0.946)	\cr
				\multirow{2}{*}[1.2ex]{\emph{R-Centralized-FC}}	&0.8301	\\ (0.829, 0.831)
														&\textcolor{red}{0.8977} \\ (0.898, 0.898)
														&0.9137	\\ (0.914, 0.914)
														&0.9265	\\ (0.926, 0.927)
														&0.9383	\\ (0.938, 0.938)
														&0.9417	\\ (0.942, 0.942)
														&\textcolor{red}{0.9463} \\ (0.946, 0.946)	\cr
				\multirow{2}{*}[1.2ex]{\emph{R-FedAvg}}	&0.0669	\\ (0.065, 0.068)
														&0.2203	\\ (0.214, 0.225)
														&0.3214	\\ (0.320, 0.328)
														&0.3250	\\ (0.320, 0.327)
														&0.3240	\\ (0.322, 0.325)
														&0.3829	\\ (0.383, 0.388)
														&0.3366 \\ (0.332, 0.337)	\cr
				\multirow{2}{*}[1.2ex]{\emph{R-FedProx}}	&0.1086	\\ (0.107, 0.115)
														&0.2110	\\ (0.208, 0.214)
														&0.3468	\\ (0.345, 0.350)
														&0.3576	\\ (0.357, 0.361)
														&0.3627	\\ (0.360, 0.366)
														&0.3468	\\ (0.346, 0.350)
														&0.3981 \\ (0.397, 0.402)	\cr
				\multirow{2}{*}[1.2ex]{\emph{R-FedPAQ}}		&0.1146	\\ (0.111, 0.116)
														&0.2527	\\ (0.252, 0.257)
														&0.2789	\\ (0.277, 0.282)
														&0.4285	\\ (0.426, 0.430)
														&0.3773	\\ (0.374, 0.378)
														&0.3930	\\ (0.392, 0.394)
														&0.4032 \\ (0.402, 0.409)	\cr
				\multirow{2}{*}[1.2ex]{\emph{R-FedAvg-FC}}   	&0.7912	\\ (0.791, 0.792)
														&\textbf{0.8527} \\ (0.852, 0.853)
														&0.8610	\\ (0.860, 0.861)
														&0.8805	\\ (0.880, 0.881)
														&0.8896	\\ (0.889, 0.890)
														&0.8841	\\ (0.884, 0.884)
														&0.8971 \\ (0.897, 0.898)	\cr
				\multirow{2}{*}[1.2ex]{\emph{R-FedDBL(ours)}} 	&\textbf{\textcolor{red}{0.8833}} \\ (0.883, 0.884)
																&0.8502	\\ (0.850, 0.851)
																&\textbf{\textcolor{red}{0.9268}} \\ (0.926, 0.927)
																&\textbf{\textcolor{red}{0.9443}} \\ (0.944, 0.945)
																&\textbf{\textcolor{red}{0.9443}} \\ (0.944, 0.945)
																&\textbf{\textcolor{red}{0.9445}} \\ (0.944, 0.945)
																&\textbf{0.9432} \\ (0.943, 0.944)	\cr
				
				\cdashline{1-8}
				\multirow{2}{*}[1.2ex]{\emph{E-Centralized}} 	&0.7956 \\ (0.794, 0.798)
																&0.8999 \\ (0.899, 0.901)
																&0.9189 \\ (0.917, 0.919)
																&\textcolor{red}{0.9465} \\ (0.946, 0.947)
																&\textcolor{red}{0.9464} \\ (0.944, 0.948)
																&\textcolor{red}{0.9589} \\ (0.959, 0.959)
																&\textcolor{red}{0.9614} \\ (0.961, 0.962)	\cr
				\multirow{2}{*}[1.2ex]{\emph{E-Centralized-FC}}		&0.8079 \\ (0.807, 0.808)
																	&0.8793 \\ (0.879, 0.880)
																	&0.9008 \\ (0.900, 0.901)
																	&0.9249 \\ (0.924, 0.925)
																	&0.9323 \\ (0.932, 0.933)
																	&0.9349 \\ (0.934, 0.935)
																	&0.9376 \\ (0.937, 0.938)	\cr
				\multirow{2}{*}[1.2ex]{\emph{E-FedAvg}}     &0.6790 \\ (0.675, 0.684)
															&0.7902 \\ (0.789, 0.791)
															&0.7914 \\ (0.791, 0.794)
															&0.8199 \\ (0.820, 0.823)
															&0.8122 \\ (0.810, 0.814)
															&0.8155 \\ (0.815, 0.816)
															&0.8222 \\ (0.821, 0.823)	\cr
				\multirow{2}{*}[1.2ex]{\emph{E-FedProx}}    &0.7112 \\ (0.709, 0.714)
															&0.7852 \\ (0.780, 0.787)
															&0.7667 \\ (0.764, 0.767)
															&0.7973 \\ (0.795, 0.799)
															&0.8207 \\ (0.819, 0.821)
															&0.8088 \\ (0.808, 0.812)
															&0.8209 \\ (0.820, 0.822)\cr
				\multirow{2}{*}[1.2ex]{\emph{E-FedPAQ}}		&0.6623 \\ (0.658, 0.665)
															&0.7848 \\ (0.782, 0.787)
															&0.7821 \\ (0.780, 0.787)
															&0.8090 \\ (0.807, 0.809)
															&0.8384 \\ (0.837, 0.840)
															&0.8275 \\ (0.827, 0.831)
															&0.8200 \\ (0.819, 0.821)	\cr
				\multirow{2}{*}[1.2ex]{\emph{E-FedAvg-FC}}   	&0.7045 \\ (0.703, 0.706)
																&0.8288 \\ (0.828, 0.829)
																&0.8522 \\ (0.852, 0.853)
																&0.8814 \\ (0.881, 0.881)
																&0.8907 \\ (0.890, 0.891)
																&0.8973 \\ (0.897, 0.898)
																&0.9008 \\ (0.900, 0.901)	\cr
				\multirow{2}{*}[1.2ex]{\emph{E-FedDBL(ours)}}    &\textbf{\textcolor{red}{0.8525}} \\ (0.852, 0.853)
																	&\textbf{\textcolor{red}{0.9159}} \\ (0.915, 0.916)
																	&\textbf{\textcolor{red}{0.9298}} \\ (0.929, 0.930)
																	&\textbf{0.9348} \\ (0.934, 0.935)
																	&\textbf{0.9345} \\ (0.934, 0.935)
																	&\textbf{0.9345} \\ (0.934, 0.935)
																	&\textbf{0.9347} \\ (0.934, 0.935)	\cr

				\cdashline{1-8}
				\multirow{2}{*}[1.2ex]{\emph{C-Centralized-FC}}		&\textcolor{red}{0.9396} \\ (0.939, 0.940)
															&0.9599	\\ (0.959, 0.96)
															&\textcolor{red}{0.9673} \\ (0.967, 0.968)
															&\textcolor{red}{0.9757} \\ (0.975, 0.976)
															&\textcolor{red}{0.9789} \\ (0.978, 0.979)
															&\textcolor{red}{0.9802} \\ (0.979, 0.980)
															&\textcolor{red}{0.9817} \\ (0.981, 0.982)	\cr
				\multirow{2}{*}[1.2ex]{\emph{C-FedAvg-FC}} 		&0.9031	\\ (0.903, 0.904)
														&0.9385 \\ (0.938, 0.939)
														&0.9459	\\ (0.945, 0.946)
														&0.9541	\\ (0.954, 0.954)
														&0.9568	\\ (0.956, 0.957)
														&0.9582	\\ (0.958, 0.959)
														&0.9600	\\ (0.959, 0.960)	\cr
				\multirow{2}{*}[1.2ex]{\emph{C-FedDBL(ours)}} &\textbf{0.9217} \\ (0.921, 0.922)
														&\textbf{\textcolor{red}{0.9643}} \\ (0.964, 0.964)
														&\textbf{0.9657} \\ (0.965, 0.966)
														&\textbf{0.9666} \\ (0.966, 0.967)
														&\textbf{0.9671} \\ (0.967, 0.967)
														&\textbf{0.9672} \\ (0.967, 0.967)
														&\textbf{0.9672} \\ (0.967, 0.967)	\cr
				\bottomrule
				\hspace{1mm}
		\end{tabular}}
	\end{threeparttable}
	\label{tab:CRC_one-round-f1-score}
\end{table*}

\begin{table*}[th]
	\centering
	\caption{Average F1-score on BCSS dataset with different proportions of the training data under one-round training. For simplification, the detail settings can be found in Table~\ref{tab:CRC_one-round-mcc}.}
	\begin{threeparttable}
		\setlength{\tabcolsep}{2.5mm}{
			\begin{tabular}{p{0.09\textwidth}>{\centering}
					p{0.09\textwidth}>{\centering}
					p{0.09\textwidth}>{\centering}
					p{0.09\textwidth}>{\centering}
					p{0.09\textwidth}>{\centering}
					p{0.09\textwidth}>{\centering}
					p{0.09\textwidth}>{\centering}
					p{0.09\textwidth}}
				\toprule
				\multirow{2}{*}{Models} & \multicolumn{7}{c}{F1-score under One-round Training on BCSS} \cr 
				\cmidrule(lr){2-8} 
				&1\% &5\% &10\% &30\% &50\% &70\% &100\%  \cr
				\midrule
				\multirow{2}{*}[1.2ex]{\emph{R-Centralized}}	&0.5851 \\ (0.569, 0.587)
														&0.2486 \\ (0.244, 0.253)
														&0.4529 \\ (0.448, 0.466)
														&0.8262 \\ (0.826, 0.840)
														&0.8255 \\ (0.824, 0.829)
														&0.8335 \\ (0.825, 0.837)
														&0.8955 \\ (0.893, 0.897)	\cr
				\multirow{2}{*}[1.2ex]{\emph{R-Centralized-FC}}	&0.2056	\\ (0.204, 0.208)
															&0.4846	\\ (0.483, 0.486)
															&0.7297	\\ (0.726, 0.732)
															&0.8979	\\ (0.896, 0.898)
															&0.9231	\\ (0.922, 0.923)
															&0.9304	\\ (0.930, 0.931)
															&0.9404 \\ (0.939, 0.941)	\cr
				\multirow{2}{*}[1.2ex]{\emph{R-FedAvg}}	&0.2300 \\ (0.225, 0.231)
													&0.4059 \\ (0.405, 0.408)
													&0.3689 \\ (0.368, 0.375)
													&0.3433 \\ (0.336, 0.346)
													&0.2962 \\ (0.293, 0.305)
													&0.2176 \\ (0.215, 0.229)
													&0.2798 \\ (0.276, 0.287)	\cr
				\multirow{2}{*}[1.2ex]{\emph{R-FedProx}}	&0.2276 \\ (0.223, 0.229)
													&0.3968 \\ (0.393, 0.398)
													&0.3578 \\ (0.350, 0.362)
													&0.3600 \\ (0.354, 0.363)
													&0.3024 \\ (0.299, 0.311)
													&0.3052 \\ (0.298, 0.307)
													&0.2633 \\ (0.254, 0.265)	\cr
				\multirow{2}{*}[1.2ex]{\emph{R-FedPAQ}}	&0.2207	\\ (0.215, 0.221)
													&0.3096	\\ (0.301, 0.313)
													&0.2979	\\ (0.289, 0.302)
													&0.3351	\\ (0.334, 0.339)
													&0.2420	\\ (0.239, 0.244)
													&0.3305	\\ (0.323, 0.333)
													&0.2419 \\ (0.241, 0.247)	\cr
				\multirow{2}{*}[1.2ex]{\emph{R-FedAvg-FC}}	&0.2004 \\ (0.199, 0.202)
													&0.2650 \\ (0.258, 0.266)
													&0.4161 \\ (0.415, 0.417)
													&0.7445 \\ (0.742, 0.747)
													&0.8192 \\ (0.819, 0.820)
													&0.8764 \\ (0.876, 0.877)
													&0.8959 \\ (0.895, 0.896)	\cr
				\multirow{2}{*}[1.2ex]{\emph{R-FedDBL(ours)}}	&\textbf{\textcolor{red}{0.8471}} \\ (0.845, 0.848)
															&\textbf{\textcolor{red}{0.9227}} \\ (0.921, 0.923)
															&\textbf{\textcolor{red}{0.9413}} \\ (0.941, 0.941)
															&\textbf{\textcolor{red}{0.9578}} \\ (0.957, 0.958)
															&\textbf{\textcolor{red}{0.9638}} \\ (0.964, 0.964)
															&\textbf{\textcolor{red}{0.9621}} \\ (0.962, 0.962)
															&\textbf{\textcolor{red}{0.9550}} \\ (0.954, 0.955)	\cr
				
				\cdashline{1-8}
				\multirow{2}{*}[1.2ex]{\emph{E-Centralized}} 		&0.4008 \\ (0.395, 0.403)
																	&0.7595 \\ (0.753, 0.762)
																	&0.8533 \\ (0.851, 0.855)
																	&0.8273 \\ (0.824, 0.832)
																	&0.8922 \\ (0.888, 0.894)
																	&0.8996 \\ (0.895, 0.900)
																	&0.9230 \\ (0.921, 0.924)	\cr
				\multirow{2}{*}[1.2ex]{\emph{E-Centralized-FC}}     &0.2985 \\ (0.293, 0.300)
																	&0.2632 \\ (0.260, 0.264)
																	&0.4275 \\ (0.427, 0.429)
																	&0.7120 \\ (0.709, 0.713)
																	&0.8190 \\ (0.818, 0.819)
																	&0.8830 \\ (0.882, 0.883)
																	&0.9062 \\ (0.905, 0.906)	\cr
				\multirow{2}{*}[1.2ex]{\emph{E-FedAvg}}     &0.1760 \\ (0.173, 0.181)
															&0.4872 \\ (0.479, 0.492)
															&0.6311 \\ (0.627, 0.640)
															&0.8992 \\ (0.897, 0.900)
															&0.8834 \\ (0.880, 0.884)
															&0.9083 \\ (0.907, 0.910)
															&0.8920 \\ (0.891, 0.895)	\cr
				\multirow{2}{*}[1.2ex]{\emph{E-FedProx}}    &0.1816 \\ (0.179, 0.186)
															&0.5028 \\ (0.494, 0.506)
															&0.6901 \\ (0.689, 0.697)
															&0.8936 \\ (0.892, 0.894)
															&0.8950 \\ (0.894, 0.897)
															&0.9087 \\ (0.907, 0.909)
															&0.9038 \\ (0.903, 0.906)	\cr
				\multirow{2}{*}[1.2ex]{\emph{E-FedPAQ}}		&0.2117 \\ (0.205, 0.213)
															&0.4940 \\ (0.492, 0.499)
															&0.7115 \\ (0.704, 0.713)
															&0.8951 \\ (0.893, 0.897)
															&0.9017 \\ (0.901, 0.903)
															&0.9122 \\ (0.911, 0.913)
															&0.8970 \\ (0.896, 0.898)	\cr
				\multirow{2}{*}[1.2ex]{\emph{E-FedAvg-FC}}  	&0.2560 \\ (0.249, 0.258)
																&0.2284 \\ (0.226, 0.232)
																&0.1932 \\ (0.193, 0.193)
																&0.4306 \\ (0.430, 0.431)
																&0.4989 \\ (0.498, 0.501)
																&0.6515 \\ (0.651, 0.652)
																&0.7428 \\ (0.741, 0.743)	\cr
				\multirow{2}{*}[1.2ex]{\emph{E-FedDBL(ours)}}	&\textbf{\textcolor{red}{0.8141}} \\ (0.814, 0.815)
																	&\textbf{\textcolor{red}{0.8890}} \\ (0.888, 0.890)
																	&\textbf{\textcolor{red}{0.8949}} \\ (0.894, 0.895)
																	&\textbf{\textcolor{red}{0.9187}} \\ (0.918, 0.919)
																	&\textbf{\textcolor{red}{0.9359}} \\ (0.935, 0.936)
																	&\textbf{\textcolor{red}{0.9395}} \\ (0.939, 0.940)
																	&\textbf{\textcolor{red}{0.9430}} \\ (0.942, 0.943)	\cr

				\cdashline{1-8}
				\multirow{2}{*}[1.2ex]{\emph{C-Centralized-FC}}	&0.2572	\\ (0.255, 0.261)
															&0.2191	\\ (0.217, 0.222)
															&0.3272	\\ (0.326, 0.328)
															&0.7578	\\ (0.748, 0.750)
															&0.9035	\\ (0.903, 0.904)
															&0.9392	\\ (0.938, 0.939)
															&0.9614 \\ (0.961, 0.961)	\cr
				\multirow{2}{*}[1.2ex]{\emph{C-FedAvg-FC}} 	&0.2148	\\ (0.214, 0.216)
															&0.1981	\\ (0.197, 0.200)
															&0.1816	\\ (0.181, 0.182)
															&0.3272	\\ (0.326, 0.328)
															&0.4571	\\ (0.457, 0.458)
															&0.6165	\\ (0.616, 0.617)
															&0.8061 \\ (0.805, 0.807)	\cr
				\multirow{2}{*}[1.2ex]{\emph{C-FedDBL(ours)}}	&\textbf{\textcolor{red}{0.9416}} \\ (0.941, 0.942)
															&\textbf{\textcolor{red}{0.9644}} \\ (0.964, 0.965)
															&\textbf{\textcolor{red}{0.9685}} \\ (0.968, 0.969)
															&\textbf{\textcolor{red}{0.9676}} \\ (0.967, 0.968)
															&\textbf{\textcolor{red}{0.9758}} \\ (0.975, 0.976)
															&\textbf{\textcolor{red}{0.9818}} \\ (0.981, 0.982)
															&\textbf{\textcolor{red}{0.9859}} \\ (0.985, 0.986)	\cr
				
				\bottomrule
				\hspace{1mm}
		\end{tabular}}
	\end{threeparttable}
	\label{tab:BCSS_one-round-f1-score}
\end{table*}
\begin{table*}[th]
	\centering
	\caption{Comparisons with different methods on 50-round training (Accuracy). For simplification, the detail settings can be found in Table~\ref{tab:CRC_one-round-mcc}. $\dagger$ means models trained for 50 rounds.}
	\begin{threeparttable}
		\setlength{\tabcolsep}{2.5mm}{
			\begin{tabular}{p{0.04\textwidth}>{}
					p{0.09\textwidth}>{\centering}
					p{0.09\textwidth}>{\centering}
					p{0.09\textwidth}>{\centering}
					p{0.09\textwidth}>{\centering}
					p{0.09\textwidth}>{\centering}
					p{0.09\textwidth}>{\centering}
					p{0.09\textwidth}>{\centering}
					p{0.09\textwidth}}
				\toprule
				\multirow{2}{*}{Datasets} & \multirow{2}{*}{Models} & \multicolumn{7}{c}{Accuracy under Multiple-round Training} \cr 
				\cmidrule(lr){3-9} 
				&	&1\% &5\% &10\% &30\% &50\% &70\% &100\%  \cr
				\midrule
				\multirow{18}{*}{\makecell[l]{MC-\\CRC}}
					&\multirow{2}{*}[1.2ex]{\emph{R-Centralized$\dagger$}} 	&0.8830	\\ (0.882, 0.883)
																			&0.9308	\\ (0.930, 0.932)
																			&0.9508	\\ (0.951, 0.952)
																			&0.9713 \\ (0.971, 0.971)
																			&0.9789 \\ (0.978, 0.979)
																			&0.9817 \\ (0.981, 0.982)
																			&0.9846 \\ (0.984, 0.985)	\cr
					&\multirow{2}{*}[1.2ex]{\emph{E-Centralized$\dagger$}} 	&0.8745 \\ (0.873, 0.878)
																			&0.9422 \\ (0.941, 0.943)
																			&0.9581 \\ (0.958, 0.958)
																			&\textcolor{red}{0.9733} \\ (0.973, 0.974)
																			&\textcolor{red}{0.9795} \\ (0.979, 0.980)
																			&\textcolor{red}{0.9838} \\ (0.983, 0.984)
																			&\textcolor{red}{0.9860} \\ (0.986, 0.986)	\cr

					\cdashline{2-9}
					&\multirow{2}{*}[1.2ex]{\emph{R-FedAvg$\dagger$}}	&0.8747	\\ (0.874, 0.875)
																		&0.9289	\\ (0.928, 0.929)
																		&0.9385	\\ (0.938, 0.939)
																		&0.9542	\\ (0.954, 0.955)
																		&0.9590	\\ (0.959, 0.959)
																		&0.9604	\\ (0.960, 0.961)
																		&0.9629 \\ (0.962, 0.963)	\cr
					&\multirow{2}{*}[1.2ex]{\emph{R-FedProx$\dagger$}}    	&0.8786	\\ (0.878, 0.879)
																			&0.9289	\\ (0.928, 0.929)
																			&0.9409	\\ (0.940, 0.941)
																			&0.9549	\\ (0.954, 0.955)
																			&0.9586	\\ (0.958, 0.959)
																			&0.9615	\\ (0.961, 0.962)
																			&0.9622 \\ (0.962, 0.962)	\cr
					&\multirow{2}{*}[1.2ex]{\emph{E-FedAvg$\dagger$}}      	&0.9175 \\ (0.917, 0.918)
																			&0.9433 \\ (0.943, 0.943)
																			&0.9493 \\ (0.949, 0.950)
																			&0.9621 \\ (0.962, 0.962)
																			&0.9655 \\ (0.965, 0.966)
																			&0.9282 \\ (0.923, 0.932)
																			&0.9701 \\ (0.970, 0.970)	\cr
					&\multirow{2}{*}[1.2ex]{\emph{E-FedProx$\dagger$}}      &0.9147 \\ (0.914, 0.915)
																			&0.9434 \\ (0.943, 0.944)
																			&0.9500 \\ (0.950, 0.950)
																			&0.9616 \\ (0.961, 0.962)
																			&0.9666 \\ (0.966, 0.967)
																			&\textbf{0.9692} \\ (0.969, 0.969)
																			&\textbf{0.9710} \\ (0.971, 0.971)	\cr
				
					&\multirow{2}{*}[1.2ex]{\emph{R-FedDBL(ours)}}	&0.8832	\\ (0.883, 0.884)
																&0.8456	\\ (0.845, 0.846)
																&0.9229	\\ (0.922, 0.923)
																&0.9411	\\ (0.941, 0.941)
																&0.9410	\\ (0.941, 0.941)
																&0.9413	\\ (0.941, 0.942)
																&0.9399 \\ (0.939, 0.940)	\cr
					&\multirow{2}{*}[1.2ex]{\emph{E-FedDBL(ours)}}	&0.8515 \\ (0.851, 0.852)
																&0.9132 \\ (0.913, 0.914)
																&0.9275 \\ (0.927, 0.928)
																&0.9331 \\ (0.932, 0.933)
																&0.9329 \\ (0.932, 0.933)
																&0.9328 \\ (0.932, 0.933)
																&0.9330 \\ (0.932, 0.933)\cr
					&\multirow{2}{*}[1.2ex]{\emph{C-FedDBL(ours)}}  	&\textbf{\textcolor{red}{0.9213}} \\ (0.921, 0.922)
																&\textbf{\textcolor{red}{0.9640}} \\ (0.964, 0.964)
																&\textbf{\textcolor{red}{0.9654}} \\ (0.965, 0.966)
																&\textbf{0.9663} \\ (0.966, 0.967)
																&\textbf{0.9668} \\ (0.966, 0.967)
																&0.9669 \\ (0.966, 0.967)
																&0.9669 \\ (0.966, 0.967)	\cr
				
				\midrule
				\multirow{18}{*}{\centering BCSS}
					&\multirow{2}{*}[1.2ex]{\emph{R-Centralized$\dagger$}}	&0.8291	\\ (0.822, 0.833)	
																			&0.9171	\\ (0.915, 0.918)
																			&0.9458	\\ (0.945, 0.947)
																			&0.9650	\\ (0.965, 0.966)
																			&0.9810	\\ (0.981, 0.981)
																			&0.9818	\\ (0.982, 0.982)
																			&0.9838 \\ (0.983, 0.984)	\cr
					&\multirow{2}{*}[1.2ex]{\emph{E-Centralized$\dagger$}} 	&0.8603 \\ (0.856, 0.861)
																			&0.9525 \\ (0.952, 0.953)
																			&0.9542 \\ (0.953, 0.955)
																			&0.9634 \\ (0.962, 0.964)
																			&0.9802 \\ (0.980, 0.981)
																			&0.9830 \\ (0.983, 0.983)
																			&0.9828 \\ (0.983, 0.984)	\cr
					\cdashline{2-9}

					&\multirow{2}{*}[1.2ex]{\emph{R-FedAvg$\dagger$}}	&0.8657	\\ (0.862, 0.868)
																		&0.9334	\\ (0.932, 0.934)
																		&0.9393	\\ (0.938, 0.940)
																		&0.9634	\\ (0.963, 0.964)
																		&0.9730	\\ (0.973, 0.973)
																		&0.9804	\\ (0.980, 0.980)
																		&0.9834 \\ (0.983, 0.983)	\cr
					&\multirow{2}{*}[1.2ex]{\emph{R-FedProx$\dagger$}}	&0.8700	\\ (0.868, 0.873)
																		&0.9201	\\ (0.918, 0.921)
																		&0.9502	\\ (0.948, 0.951)
																		&0.9671	\\ (0.967, 0.967)
																		&0.9737	\\ (0.973, 0.974)
																		&0.9776	\\ (0.977, 0.978)
																		&0.9830 \\ (0.983, 0.983)	\cr
					
					&\multirow{2}{*}[1.2ex]{\emph{E-FedAvg$\dagger$}}   &0.8608 \\ (0.859, 0.862)
																		&0.9412 \\ (0.941, 0.942)
																		&0.9451 \\ (0.943, 0.945)
																		&\textbf{\textcolor{red}{0.9782}} \\ (0.978, 0.978)
																		&0.9815 \\ (0.981, 0.982)
																		&0.9847 \\ (0.984, 0.985)
																		&0.9840 \\ (0.984, 0.984)	\cr
					&\multirow{2}{*}[1.2ex]{\emph{E-FedProx$\dagger$}}  &0.8873 \\ (0.884, 0.888)
																		&0.9395 \\ (0.938, 0.940)
																		&0.9575 \\ (0.956, 0.958)
																		&0.9763 \\ (0.976, 0.976)
																		&0.9784 \\ (0.978, 0.979)
																		&0.9849 \\ (0.984, 0.985)
																		&0.9852 \\ (0.985, 0.985)	\cr
					
					&\multirow{2}{*}[1.2ex]{\emph{R-FedDBL(ours)}}	&0.9012	\\ (0.900, 0.902)
																&0.9511	\\ (0.951, 0.951)
																&0.9603	\\ (0.960, 0.961)
																&0.9711	\\ (0.971, 0.971)
																&0.9745	\\ (0.974, 0.975)
																&0.9731	\\ (0.973, 0.973)
																&0.9652	\\ (0.965, 0.965)	\cr
					&\multirow{2}{*}[1.2ex]{\emph{E-FedDBL(ours)}}    &0.8971 \\ (0.897, 0.898)
																&0.9368 \\ (0.936, 0.937)
																&0.9324 \\ (0.932, 0.933)
																&0.9485 \\ (0.948, 0.949)
																&0.9607 \\ (0.960, 0.961)
																&0.9640 \\ (0.964, 0.964)
																&0.9656 \\ (0.965, 0.966)	\cr
					&\multirow{2}{*}[1.2ex]{\emph{C-FedDBL(ours)}}	&\textbf{\textcolor{red}{0.9593}} \\ (0.959, 0.960)
																&\textbf{\textcolor{red}{0.9754}} \\ (0.975, 0.976)
																&\textbf{\textcolor{red}{0.9777}} \\ (0.977, 0.978)
																&0.9770 \\ (0.977, 0.978)
																&\textbf{\textcolor{red}{0.9834}} \\ (0.983, 0.983)
																&\textbf{\textcolor{red}{0.9883}} \\ (0.988, 0.988)
																&\textbf{\textcolor{red}{0.9910}} \\ (0.991, 0.991)	\cr
				
				\bottomrule
				\hspace{1mm}
		\end{tabular}}
	\end{threeparttable}
	\label{tab:accuracy}
\end{table*}
\begin{table*}[th]
	\centering
	\caption{Comparisons with different methods on 50-round training (F1-score). For simplification, the detail settings can be found in Table~\ref{tab:CRC_one-round-mcc}. $\dagger$ means models trained for 50 rounds.}
	\begin{threeparttable}
		\setlength{\tabcolsep}{2.5mm}{
			\begin{tabular}{p{0.04\textwidth}>{}
					p{0.09\textwidth}>{\centering}
					p{0.09\textwidth}>{\centering}
					p{0.09\textwidth}>{\centering}
					p{0.09\textwidth}>{\centering}
					p{0.09\textwidth}>{\centering}
					p{0.09\textwidth}>{\centering}
					p{0.09\textwidth}>{\centering}
					p{0.09\textwidth}}
				\toprule
				\multirow{2}{*}{Datasets} & \multirow{2}{*}{Models} & \multicolumn{7}{c}{F1-score under Multiple-round Training} \cr 
				\cmidrule(lr){3-9} 
				&	&1\% &5\% &10\% &30\% &50\% &70\% &100\%  \cr
				\midrule
				\multirow{18}{*}{\makecell[l]{MC-\\CRC}}
					&\multirow{2}{*}[1.2ex]{\emph{R-Centralized$\dagger$}} 	&0.8813	\\ (0.881, 0.882)
																			&0.9307	\\ (0.930, 0.932)
																			&0.9512	\\ (0.951, 0.952)
																			&0.9714 \\ (0.971, 0.972)
																			&0.9789 \\ (0.978, 0.979)
																			&0.9816 \\ (0.981, 0.982)
																			&0.9845 \\ (0.984, 0.985)	\cr
					&\multirow{2}{*}[1.2ex]{\emph{E-Centralized$\dagger$}} 	&0.8741 \\ (0.872, 0.877)
																			&0.9427 \\ (0.942, 0.943)
																			&0.9581 \\ (0.958, 0.958)
																			&\textcolor{red}{0.9732} \\ (0.973, 0.973)
																			&\textcolor{red}{0.9794} \\ (0.979, 0.980)
																			&\textcolor{red}{0.9837} \\ (0.983, 0.984)
																			&\textcolor{red}{0.9859} \\ (0.985, 0.986)	\cr

					\cdashline{2-9}
					&\multirow{2}{*}[1.2ex]{\emph{R-FedAvg$\dagger$}}	&0.8720	\\ (0.872, 0.873)
																		&0.9283	\\ (0.928, 0.928)
																		&0.9386	\\ (0.938, 0.939)
																		&0.9545	\\ (0.954, 0.955)
																		&0.9592	\\ (0.959, 0.959)
																		&0.9607	\\ (0.960, 0.961)
																		&0.9632 \\ (0.963, 0.963)	\cr
					&\multirow{2}{*}[1.2ex]{\emph{R-FedProx$\dagger$}}	&0.8751	\\ (0.875, 0.876)
																		&0.9288	\\ (0.928, 0.929)
																		&0.9409	\\ (0.940, 0.941)
																		&0.9550	\\ (0.955, 0.955)
																		&0.9591	\\ (0.959, 0.959)
																		&0.9618	\\ (0.961, 0.962)
																		&0.9625 \\ (0.962, 0.963)	\cr
					
					&\multirow{2}{*}[1.2ex]{\emph{E-FedAvg$\dagger$}}      	&0.9172 \\ (0.917, 0.918)
																			&0.9434 \\ (0.943, 0.943)
																			&0.9494 \\ (0.949, 0.950)
																			&0.9623 \\ (0.962, 0.962)
																			&0.9655 \\ (0.965, 0.966)
																			&0.9206 \\ (0.914, 0.926)
																			&0.9701 \\ (0.970, 0.971)	\cr
					&\multirow{2}{*}[1.2ex]{\emph{E-FedProx$\dagger$}}      &0.9147 \\ (0.914, 0.915)
																			&0.9437 \\ (0.943, 0.944)
																			&0.9502 \\ (0.950, 0.950)
																			&0.9617 \\ (0.962, 0.962)
																			&0.9669 \\ (0.966, 0.967)
																			&\textbf{0.9693} \\ (0.969, 0.969)
																			&\textbf{0.9712} \\ (0.971, 0.972)	\cr
				
					&\multirow{2}{*}[1.2ex]{\emph{R-FedDBL(ours)}}	&0.8833	\\ (0.883, 0.884)
														&0.8502	\\ (0.850, 0.851)
														&0.9268	\\ (0.926, 0.927)
														&0.9443	\\ (0.944, 0.945)
														&0.9443	\\ (0.944, 0.945)
														&0.9445	\\ (0.944, 0.945)
														&0.9432 \\ (0.943, 0.944)	\cr
					&\multirow{2}{*}[1.2ex]{\emph{E-FedDBL(ours)}}	&0.8525 \\ (0.852, 0.853)
																&0.9159 \\ (0.915, 0.916)
																&0.9298 \\ (0.929, 0.930)
																&0.9348 \\ (0.934, 0.935)
																&0.9345 \\ (0.934, 0.935)
																&0.9345 \\ (0.934, 0.935)
																&0.9347 \\ (0.934, 0.935)	\cr
					&\multirow{2}{*}[1.2ex]{\emph{C-FedDBL(ours)}}	&\textbf{\textcolor{red}{0.9217}} \\ (0.921, 0.922)
														&\textbf{\textcolor{red}{0.9643}} \\ (0.964, 0.964)
														&\textbf{\textcolor{red}{0.9657}} \\ (0.965, 0.966)
														&\textbf{0.9666} \\ (0.966, 0.967)
														&\textbf{0.9671} \\ (0.967, 0.967)
														&0.9672 \\ (0.967, 0.967)
														&0.9672 \\ (0.967, 0.967)	\cr
				
				\midrule
				\multirow{18}{*}{\centering BCSS}
					&\multirow{2}{*}[1.2ex]{\emph{R-Centralized$\dagger$}}	&0.7671	\\ (0.760, 0.771)
																		&0.8798	\\ (0.877, 0.881)
																		&0.9182	\\ (0.918, 0.920)
																		&0.9554	\\ (0.955, 0.957)
																		&0.9722	\\ (0.972, 0.972)
																		&0.9756	\\ (0.975, 0.976)
																		&0.9761	\\ (0.975, 0.977)	\cr
					&\multirow{2}{*}[1.2ex]{\emph{E-Centralized$\dagger$}} 	&0.7782 \\ (0.772, 0.779)
																			&0.9280 \\ (0.927, 0.929)
																			&0.9370 \\ (0.935, 0.938)
																			&0.9527 \\ (0.951, 0.954)
																			&0.9717 \\ (0.971, 0.972)
																			&0.9775 \\ (0.977, 0.978)
																			&0.9766 \\ (0.976, 0.978)	\cr

					\cdashline{2-9}
					&\multirow{2}{*}[1.2ex]{\emph{R-FedAvg$\dagger$}}	&0.8088	\\ (0.804, 0.814)
																		&0.9003	\\ (0.898, 0.901)
																		&0.8999	\\ (0.897, 0.902)
																		&0.9448	\\ (0.944, 0.946)
																		&0.9598	\\ (0.959, 0.960)
																		&0.9721	\\ (0.972, 0.972)
																		&0.9741	\\ (0.974, 0.974)	\cr
					&\multirow{2}{*}[1.2ex]{\emph{R-FedProx$\dagger$}}	&0.8095	\\ (0.807, 0.813)
																		&0.8851	\\ (0.883, 0.886)
																		&0.9293	\\ (0.926, 0.930)
																		&0.9522	\\ (0.952, 0.953)
																		&0.9599	\\ (0.959, 0.961)
																		&0.9669	\\ (0.966, 0.967)	
																		&0.9769 \\ (0.976, 0.977)	\cr
					
					&\multirow{2}{*}[1.2ex]{\emph{E-FedAvg$\dagger$}}	&0.7724 \\ (0.768, 0.773)
																		&0.9051 \\ (0.904, 0.906)
																		&0.9014 \\ (0.897, 0.903)
																		&\textbf{\textcolor{red}{0.9686}} \\ (0.968, 0.969)
																		&0.9750 \\ (0.975, 0.975)
																		&0.9791 \\ (0.979, 0.979)
																		&0.9780 \\ (0.978, 0.978)	\cr
					&\multirow{2}{*}[1.2ex]{\emph{E-FedProx$\dagger$}}	&0.8270 \\ (0.822, 0.828)
																		&0.9036 \\ (0.901, 0.904)
																		&0.9382 \\ (0.937, 0.939)
																		&0.9653 \\ (0.965, 0.966)
																		&0.9696 \\ (0.969, 0.970)
																		&0.9776 \\ (0.977, 0.978)
																		&0.9778 \\ (0.977, 0.978)	\cr
					
					&\multirow{2}{*}[1.2ex]{\emph{R-FedDBL(ours)}}	&0.8471	\\ (0.845, 0.848)
																&0.9227	\\ (0.921, 0.923)
																&0.9413	\\ (0.941, 0.941)
																&0.9578	\\ (0.957, 0.958)
																&0.9638	\\ (0.964, 0.964)
																&0.9621	\\ (0.962, 0.962)
																&0.9550	\\ (0.954, 0.955)	\cr
					&\multirow{2}{*}[1.2ex]{\emph{E-FedDBL(ours)}}    &0.8141 \\ (0.814, 0.815)
																&0.8890 \\ (0.888, 0.890)
																&0.8949 \\ (0.894, 0.895)
																&0.9187 \\ (0.918, 0.919)
																&0.9359 \\ (0.935, 0.936)
																&0.9395 \\ (0.939, 0.940)
																&0.9430 \\ (0.942, 0.943)	\cr
					&\multirow{2}{*}[1.2ex]{\emph{C-FedDBL(ours)}}	&\textbf{\textcolor{red}{0.9416}}	\\ (0.941, 0.942)
																&\textbf{\textcolor{red}{0.9644}}	\\ (0.964, 0.965)
																&\textbf{\textcolor{red}{0.9685}}	\\ (0.968, 0.969)
																&0.9676	\\ (0.967, 0.968)
																&\textbf{\textcolor{red}{0.9758}}	\\ (0.975, 0.976)
																&\textbf{\textcolor{red}{0.9818}}	\\ (0.981, 0.982)
																&\textbf{\textcolor{red}{0.9859}}	\\ (0.985, 0.986)	\cr
				
				\bottomrule
				\hspace{1mm}
		\end{tabular}}
	\end{threeparttable}
	\label{tab:f1-score}
\end{table*}
\end{document}